\documentclass[12pt]{article}  %%was 11pt
\usepackage{graphicx}
\usepackage{epsfig}
\usepackage{hyperref}
\usepackage {amsmath}
\usepackage {amssymb}
\usepackage {bm}  %%%usage $\mathbf{v_2}$
\usepackage[usenames,dvipsnames]{color}%mjt-this uses familiar color names
\usepackage{xspace} 
\newcommand{\sqsn}{\mbox{$\sqrt{s_{_{NN}}}$}\xspace}

\def\lsim{\raise0.3ex\hbox{$<$\kern-0.75em\raise-1.1ex\hbox{$\sim$}}}
\def\gsim{\raise0.3ex\hbox{$>$\kern-0.75em\raise-1.1ex\hbox{$\sim$}}}

\def\mean#1{\left<#1\right>}
\def\Journal#1#2#3#4{{#1}{\bf #2} (#4) #3}
\def\EPJH{{Eur. Phys. J. H}}
\def\JPG{{J. Phys. G}}

\def\NCA{Nuovo Cimento\ }

\def\NIMA{{Nucl. Instrum. Methods A}}
\def\NPA{{Nucl. Phys. A}}
\def\NPB{{Nucl. Phys. B}}
\def\PLB{{Phys. Lett. B}}

\def\PRL{Phys. Rev. Lett.\ }
\def\PRD{{Phys. Rev. D}}
\def\PRC{{Phys. Rev. C}}

\def\ARNPS{{Ann. Rev. Nucl. Part. Sci.\ }}

\def\la{\left< }
\def\ra{\right> }
\def\jt#1{\ensuremath{j_{T\rm #1}}}
\def\meankv#1{\ensuremath{\la#1^2\ra}}
\def\rms#1{\meankv{#1}}

\def\QGP{{\color{Red} Q}{\color{Blue} G}{\color{Green} P}} 
\def\QCD{{\color{Red} Q}{\color{Green} C}{\color{Blue} D}} 
\normalsize
\begin{document}
%%\markboth{M.~J.~Tannenbaum}{Highlights from Highlights from BNL and RHIC 2014}
\title{Highlights from BNL and RHIC 2017}
\author{M.~J.~Tannenbaum
\thanks{Research supported by U.~S.~Department of Energy, DE-SC0012704.}
\\Physics Department, 510c,\\
Brookhaven National Laboratory,\\
Upton, NY 11973-5000, USA\\
mjt@bnl.gov} 
\date{}
\maketitle
%\thispagestyle{empty} %restore  \thispagestyle{empty} for submission
%\tableofcontents
%\vspace*{-2pc}
\section{Introduction}\label{sec:introduction}
\begin{figure}[!h]
\begin{center}
\raisebox{0pc}{\includegraphics[width=0.44\textwidth]{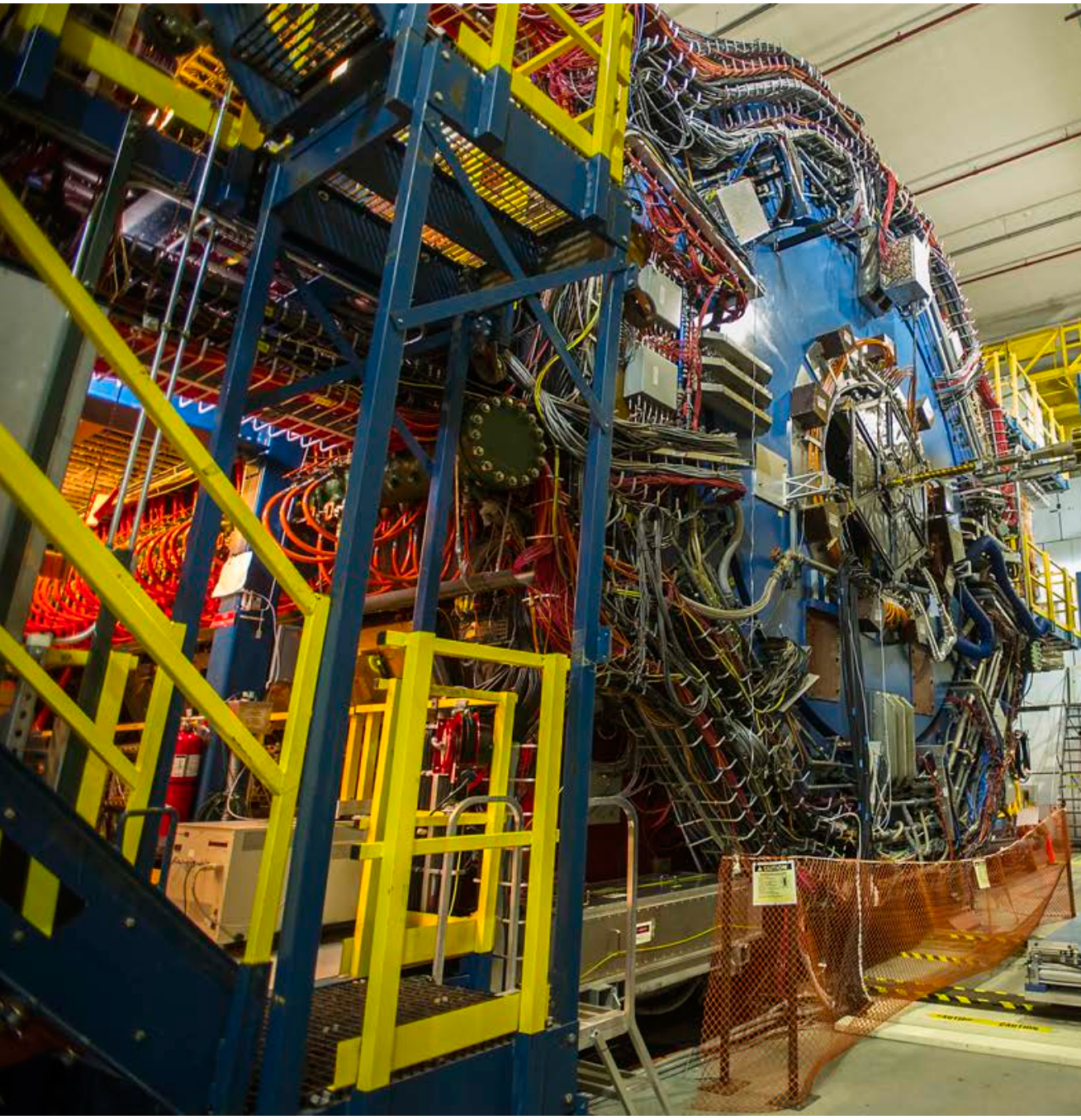}}\hspace*{0.2pc}
\raisebox{0.5pc}{\includegraphics[width=0.56\textwidth]{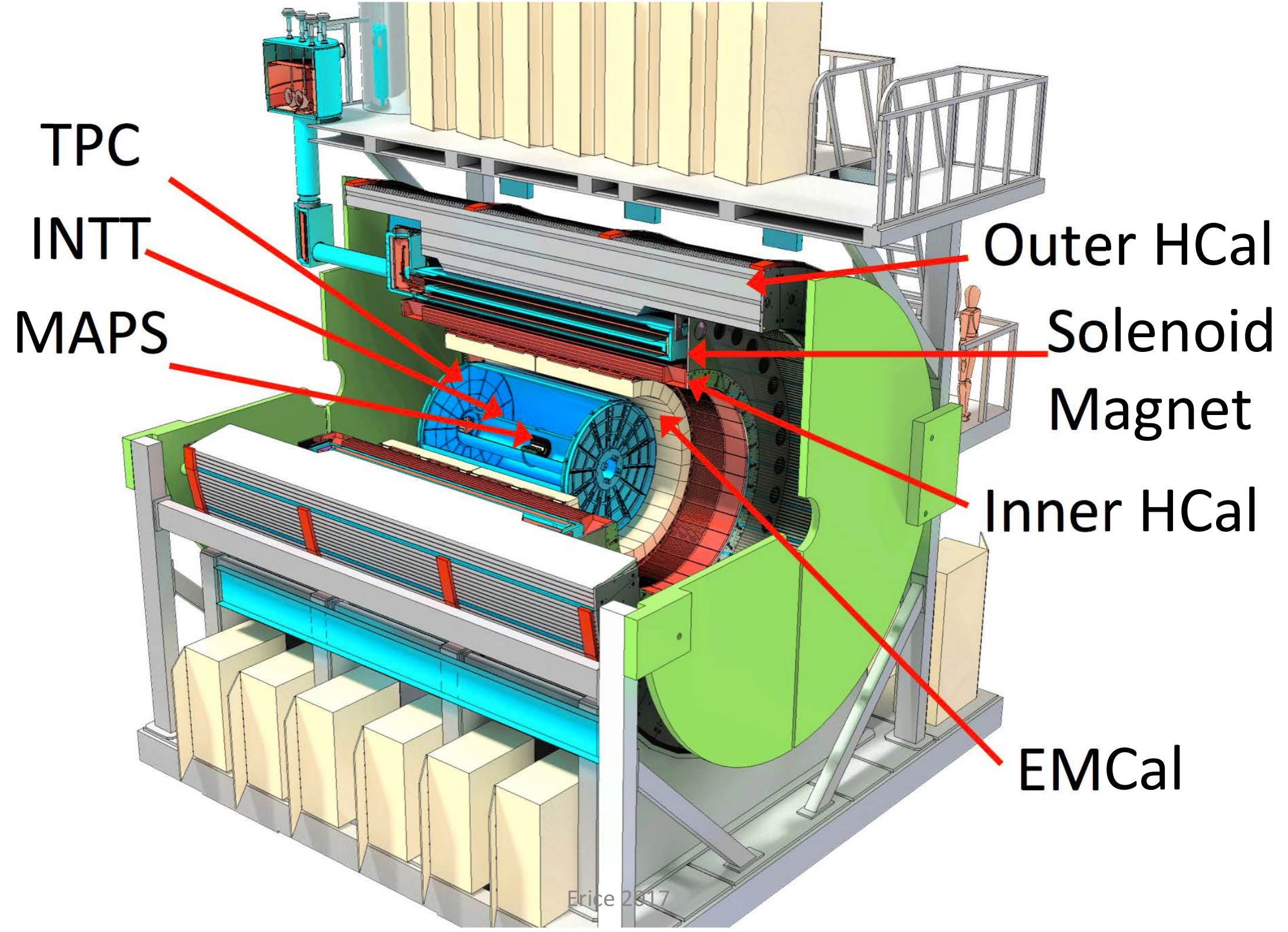}}
\end{center}\vspace*{-1.5pc}
\caption[]{\footnotesize (left) STAR experiment and (right) proposed sPHENIX. }
\label{fig:sPHENIX}\vspace*{-0.5pc}
\end{figure}

The Relativistic Heavy Ion Collider (RHIC) is one of the two remaining operating hadron colliders in the world, and the first and only polarized p$+$p collider. STAR is now the only experiment operating at RHIC. PHENIX ended its operation with the 2016 run and is now being dismantled. In its place a new experiment, sPHENIX, with a superconducting solenoid, a Hadron Calorimeter in the return yoke, an EMCal and small Hadron Calorimeter in the magnetic field as well as a Time Projection chamber for charged particle reconstruction and a silicon vertex detector (MAPS) (Fig.~\ref{fig:sPHENIX}). sPHENIX goals are precision measurements of jets, heavy flavors and Upsilon spectroscopy.

\section{Camp Upton 100 years; BNL 70, 1947-2017.}
\begin{figure}[!h]
\begin{center}
\raisebox{0pc}{\includegraphics[width=0.95\textwidth]{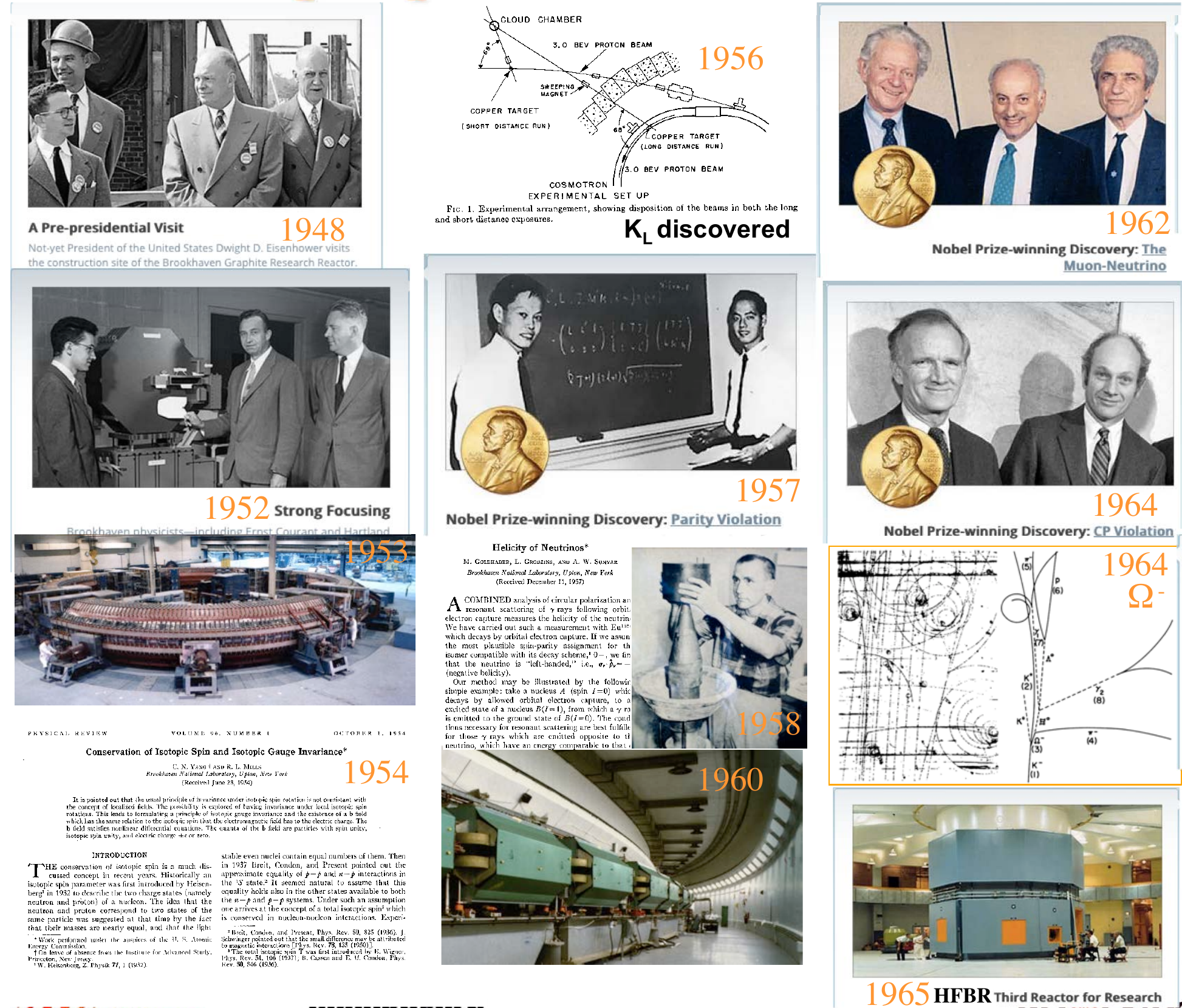}}\hspace*{0.2pc}
\end{center}\vspace*{-1.5pc}
\caption[]{\footnotesize Highlights in BNL History 1948--1965, starting with the visit of Dwight D. Eisenhower, then the president of Columbia University, in 1948. The other figures are labeled except perhaps for 1953-the Cosmotron, 1954-the Yang-Mills paper, 1958-Goldhaber, Grodzins, Sunyar, Helicity of Neutrinos.}
\label{fig:70-1}\vspace*{-0.5pc}
\end{figure}
\begin{figure}[h]
\begin{center}
\raisebox{0pc}{\includegraphics[width=0.95\textwidth]{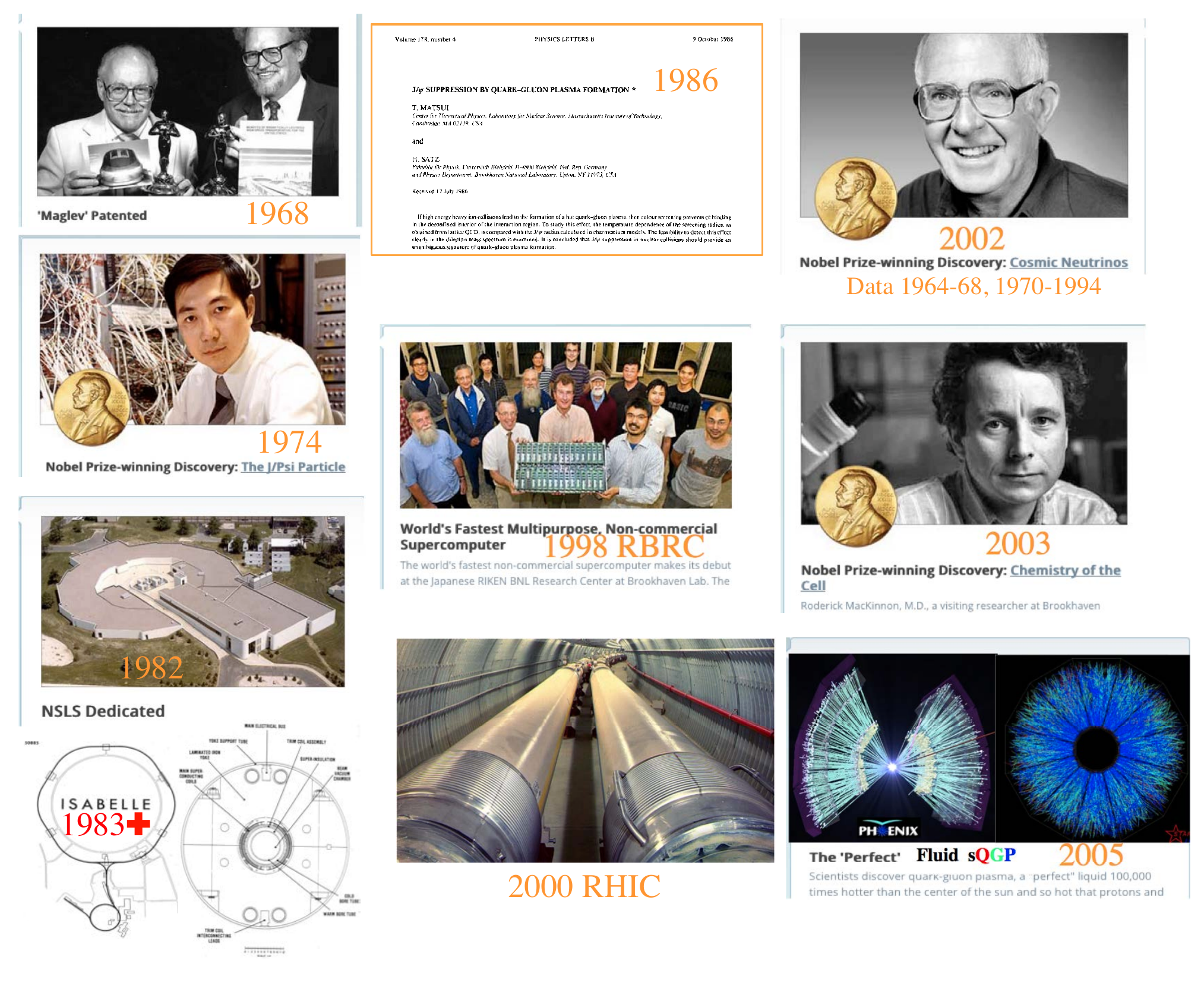}}\hspace*{0.2pc}
\end{center}\vspace*{-2.5pc}
\caption[]{\footnotesize Highlights in BNL History 1968--2005 starting with the Patent of Maglev trains by Gordon Danby and James Powell. Next to the ISABELLE p$+$p collider, cancelled in 1983, is the profile of the Palmer Magnet~\cite{PalmerMagnet}, the basis of all subsequent superconducting collider magnets. 1986-Matsui Satz $J/\Psi$ suppression, a signature of the Quark Gluon Plasma (\QGP).}
\label{fig:70-2}\vspace*{-0.5pc}
\end{figure}
In 1917, the U.S. Army opened a training camp for soldiers before they were sent off to Europe to fight in World War I. This was Camp Upton near Yaphank on Long Island, 66 miles east of Manhattan. In World War II, Camp Upton was expanded and converted into a convalescent and rehabilitation hospital in September of 1944, and was put on surplus after the war ended. In 1947, the Associated Universities decided to put their proposed laboratory for constructing, and operating large scientific machines beyond the capabilities of single universities at Camp Upton, now BNL, starting with a nuclear reactor for research studies. In addition to the scientific achievements of BNL over the past 70 years, it is interesting to note that the famous U.S. composer Irving Berlin wrote many songs while he was at Camp Upton in 1917-18, which are still popular, most notably `God Bless America', which is still played in the `7th inning stretch' at Major League baseball games. Some achievements and discoveries at BNL are pictured in Fig.~\ref{fig:70-1} for 1948-1965 and Fig.~\ref{fig:70-2} for 1968-2005.
\section{Muon g-2 experiment (from BNL) starts at Fermilab}\vspace*{-1.0pc}
\begin{minipage}[b]{0.40\linewidth}
{\includegraphics[width=0.98\textwidth]{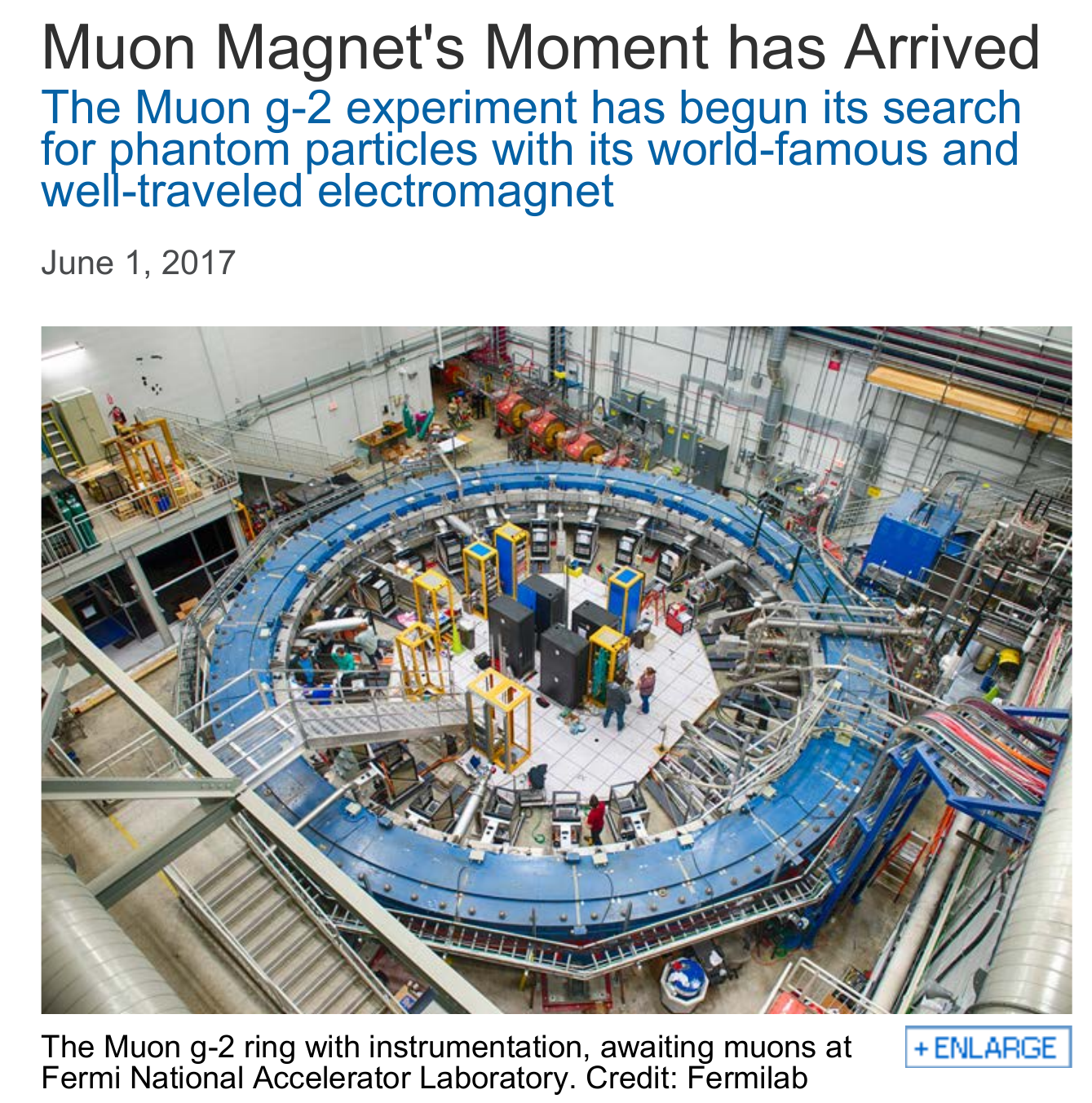}}
\end{minipage}
\begin{minipage}[b]{0.60\linewidth}
{\bf From the Fermilab press release}:
Getting to this point was a long road for Muon g-2, both figuratively and literally. The first generation of this experiment took place at {\it the U.S. DOE's Brookhaven National Laboratory in New York State in the late 1990's and early 2000's.} 

Since it would have cost 10 times more to build a completly new machine at Brookhaven rather than
move the magnet to Fermilab, the Muon g-2 team transported that large, fragile superconducting magnet in one piece from Long
Island to the suburbs of Chicago inthe summer of 2013. 
\vspace*{1.0pc}
\end{minipage} 

As usual, press releases never get it right, perhaps the person who wrote the press release never heard of 
Charpak, Farley, Garwin, Muller, Sens and Zichichi~\cite{Firstg-2}.
\subsection{Vorticity---Worthy of a Press Release}
\begin{figure}[!h]
\begin{center}
\raisebox{3.0pc}{{\footnotesize a)}\hspace*{-0.0pc}\includegraphics[width=0.52\textwidth]{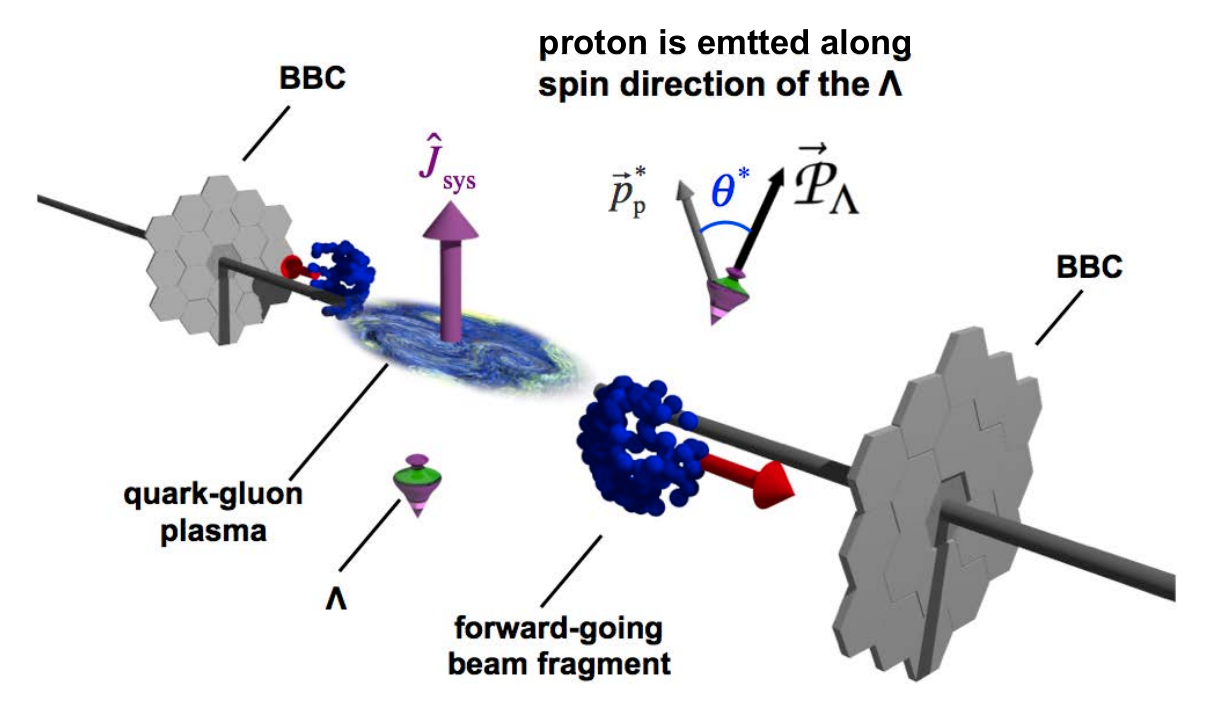}}\hspace*{0.2pc}
\raisebox{0.0pc}{{\footnotesize b)}\hspace*{-0.0pc}\includegraphics[width=0.44\textwidth]{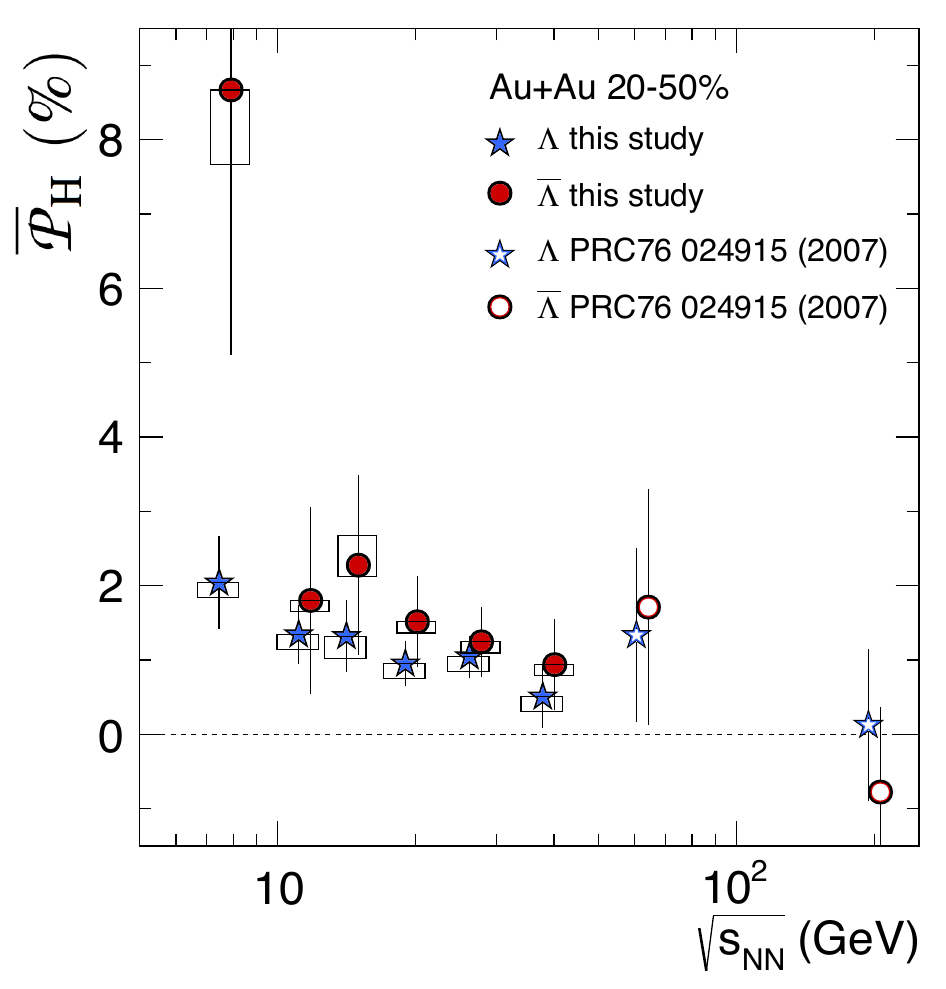}}
\end{center}\vspace*{-1.5pc}
\caption[]{\footnotesize a) Schematic of forward and backward beam fragments passing each other and producing a magnetic field with direction $\hat{j}_{\rm sys}$. b) Measured polarization ${\cal P}_{\bm H}$ with  ${\bm H}=\Lambda$ or $\bar{\Lambda}$ as a function of \sqsn.  }
\label{fig:Vorticity}\vspace*{-0.5pc}
\end{figure}
One of the most interesting new results this year from RHIC~\cite{Nature548} is a determination of the vorticity of the \QGP\ in Au$+$Au collisions by measurement of the polarization ${\cal P}_{\bm H}$ of $\Lambda$ hyperons with respect to the perpendicular of the reaction plane, which is the direction of the strong magnetic field ($\hat{j}_{\rm sys}$) formed by the current loop of the highly charged nuclei passing each other (Fig.~\ref{fig:Vorticity}). The polarization is measurable because the $\Lambda$ are generally produced polarized~\cite{BuncePRL36} and the proton in the decay $\Lambda\rightarrow p + \pi^-$ is emitted along the spin direction of the $\Lambda$. The average polarization of the $\Lambda$ and $\bar{\Lambda}$ over $7\leq\sqsn\leq200$ GeV is $\bar{\cal P}\approx (1.2\pm 0.2) \%$ from which the vorticity $\omega\approx10^{22}$/s, which is $10^{15}$ times larger than any other fluid. On the other hand it is most interesting to note that the vorticity $\rightarrow 0$ at $\sqsn=200$ GeV, where the \QGP\ ``the perfect liquid'' was discovered, and increases to its largest values in the range $\sqsn=7-19$ GeV of the CERN fixed target measurements---does this mean that they have an even more ``perfect liquid''?????

There are two other issues here that might be of interest to students: 1) calculate $\omega$ from the formula $\omega=kT\bar{\cal P}/\hbar$, where $k$ is Boltzman's constant, $\hbar$ is Plank's constant and $T$ is the temperature of the \QGP$\approx 300$ MeV; 2) See CERN 86-07 for T.D. Lee's story of how Jack Steinberger missed discovering parity violation in $\Lambda$ decay from the reaction $\pi^-+ p\rightarrow K^0 + \Lambda$.
\section{RHIC operation in 2016 and future plans}
\label{sec:RHICops}
\begin{figure}[!h]
\begin{center}
\raisebox{0pc}{\includegraphics[width=0.75\textwidth]{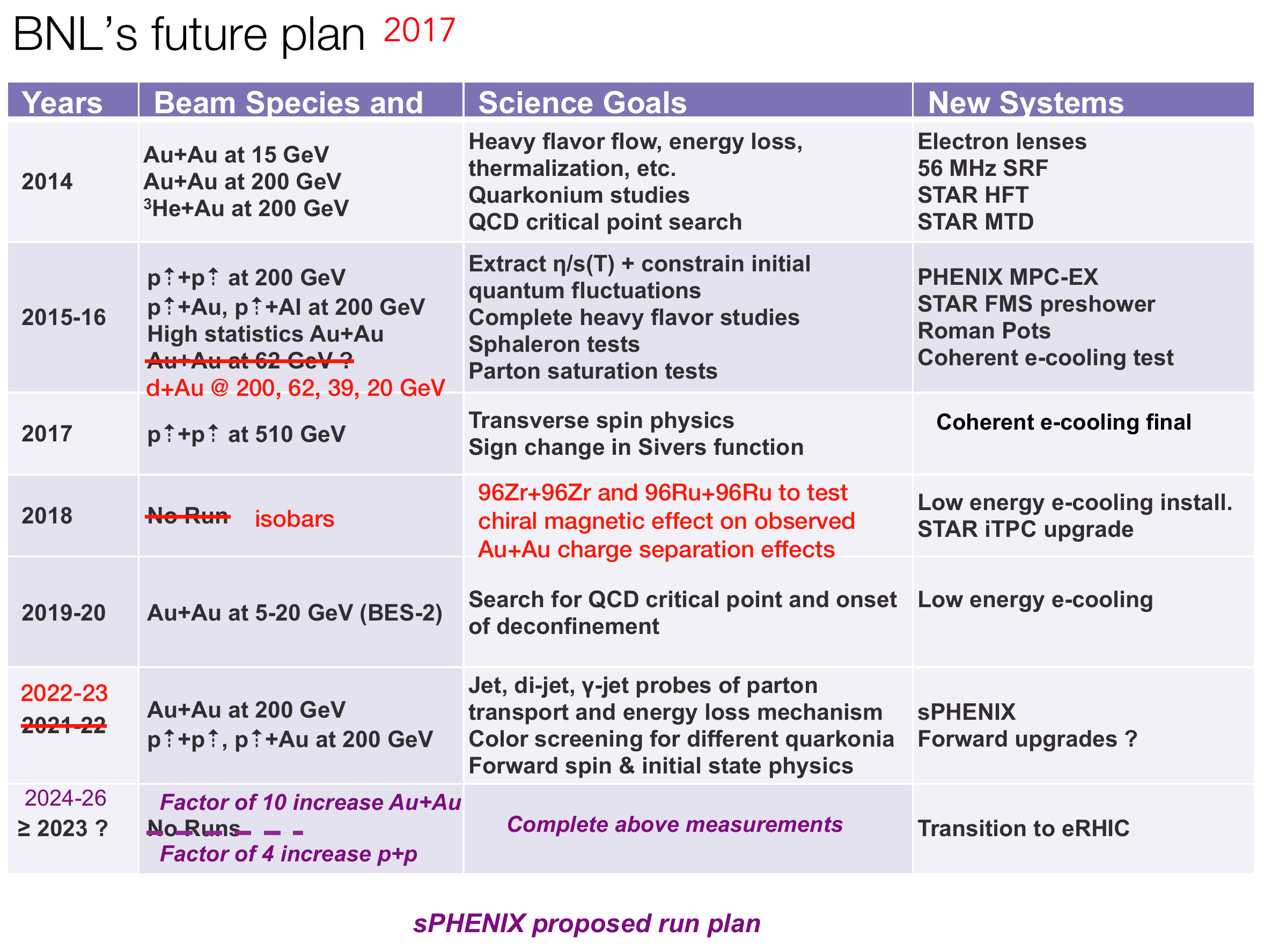}}\hspace*{0.2pc}
\end{center}\vspace*{-1.5pc}
\caption[]{\footnotesize BNL-RHIC run plan 2014--2026 }
\label{fig:RunPlan}\vspace*{-0.5pc}
\end{figure}

BNL's future plans for RHIC operation are given in Fig.~\ref{fig:RunPlan}. The main objectives until sPHENIX is working in $\approx 2022$ is a run in 2018 with collisions of isobars, $^{96}_{40}$Zr + $^{96}_{40}$Zr compared to $^{96}_{44}$Ru + $^{96}_{44}$Ru, to understand whether the charge separation of anisotropic flow $v_2$ of $\pi^+$ and $\pi^-$ observed by STAR in Au$+$Au~\cite{STARPRL114}, the so-called Chiral Magnetic Effect, will be different for the different $Z$, hence due to the strong electromagnetic field in the nuclear collisions, or will remain unchanged for collisions of nuclei with the same number of nucleons. In 2019 and 2020, STAR will then perform a beam energy scan searching for the \QGP\ critical point and onset of deconfinement.    

\begin{figure}[!h]
\begin{center}
\raisebox{0.0pc}{\includegraphics[width=0.38\textwidth]{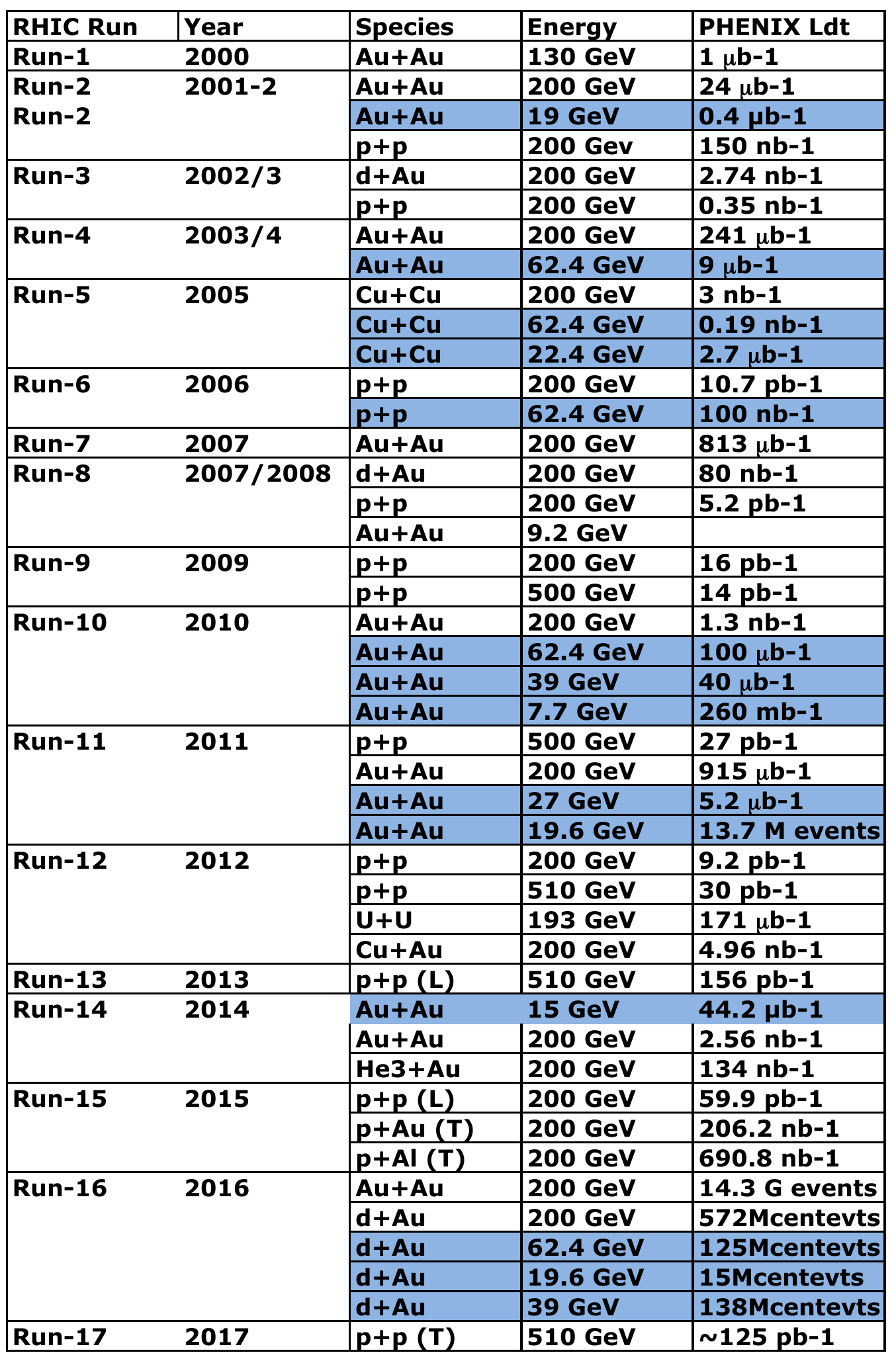}}\hspace*{0.2pc}
\raisebox{1.0pc}{\includegraphics[width=0.61\textwidth]{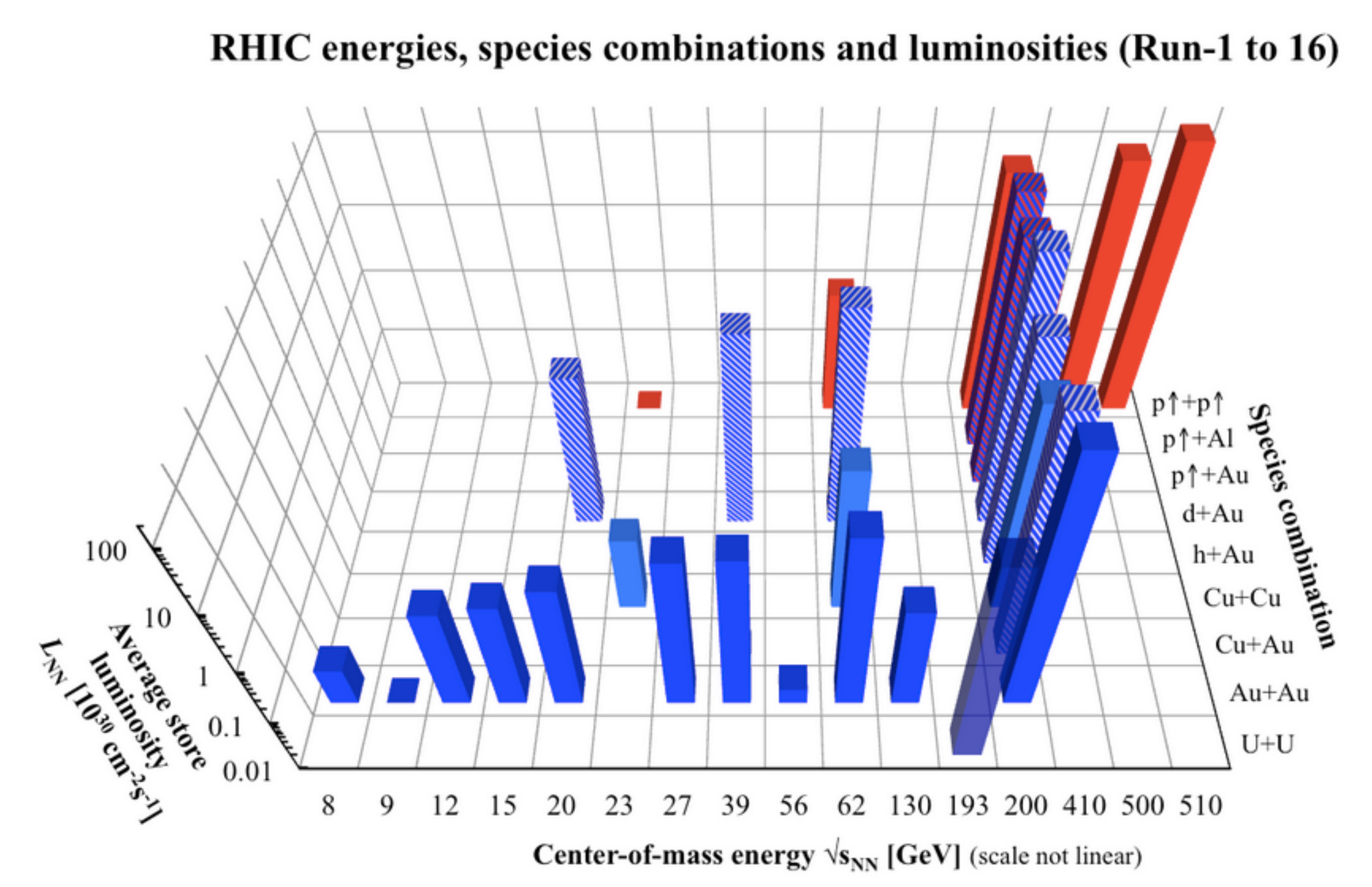}}
\end{center}\vspace*{-1.5pc}
\caption[]{\footnotesize RHICruns. }
\label{fig:RHICruns}\vspace*{-0.5pc}
\end{figure}
The history of RHIC runs is shown in Fig.~\ref{fig:RHICruns}. One may ask why the 2017 RHIC run is a transverely polarized proton run when the original idea for spin physics at RHIC was based on single spin longitudinal (parity violating) asymmetries $A_L$ of $W^{\pm}$ production, since the $W$ is coupled to flavor, not color like \QCD.
\subsection{The RHIC Spin Collaboration, Polarized Protons at RHIC}
\begin{figure}[!h]
\begin{center}
\raisebox{0.0pc}{{\footnotesize a)}\hspace*{-0.0pc}\includegraphics[width=0.7\textwidth]{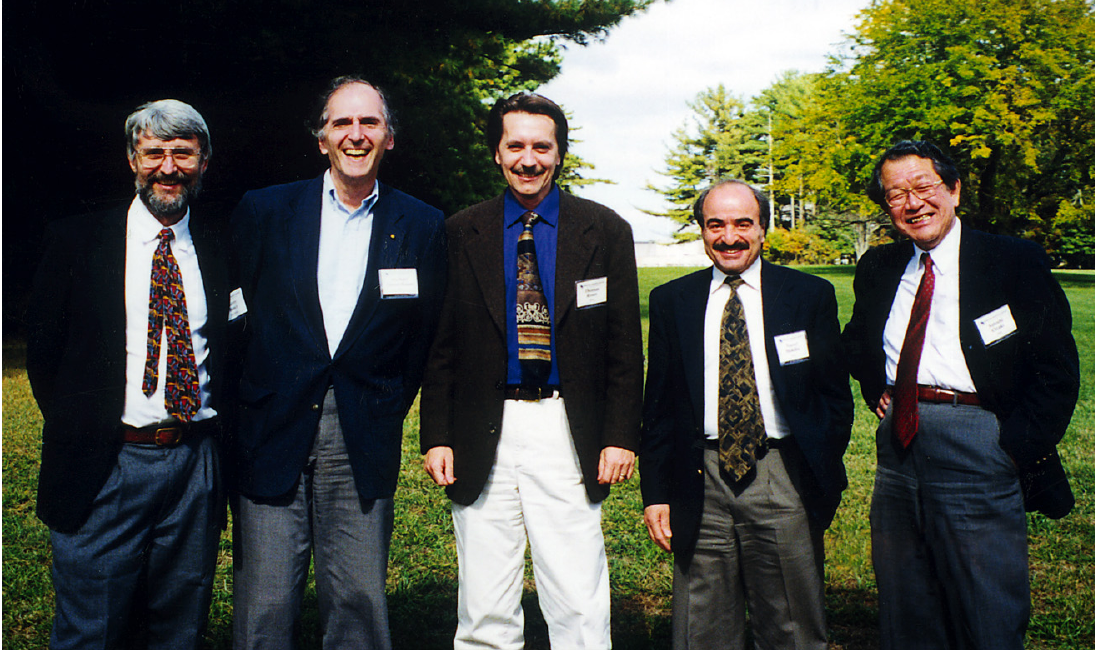}}\hspace*{0.2pc}
\raisebox{0.2pc}{{\footnotesize b)}\hspace*{-0.0pc}\includegraphics[width=0.2\textwidth]{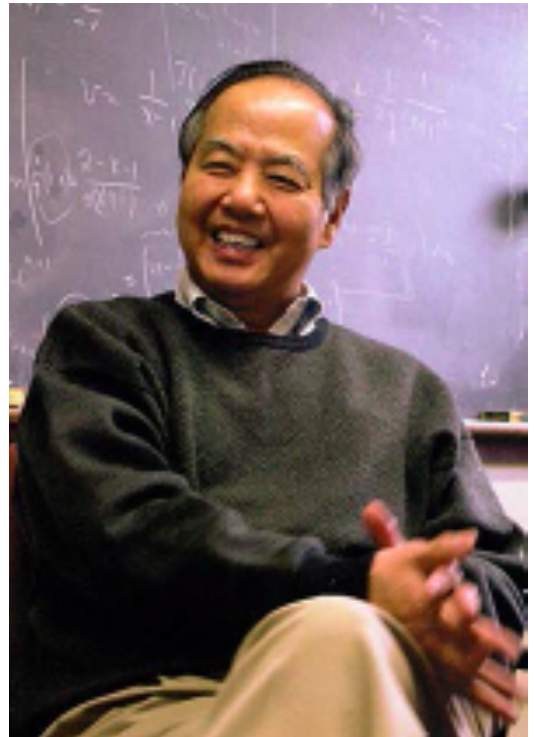}}
\end{center}\vspace*{-1.5pc}
\caption[]{\footnotesize a) BNL RSC originators (L--R) Gerry Bunce, Mike Tannenbaum, Thomas Roser, Yousef Makdisi, Satoshi Ozaki. b) T.~D.~Lee, founding director of the RIKEN BNL Research Center (RBRC). }
\label{fig:RSCoriginators}\vspace*{-0.5pc}
\end{figure}

Ideas for a polarized proton collider (then ISABELLE) started at BNL during the famous Snowmass 1982 meeting~\cite{MJTWW} and continued, when ISABELLE became RHIC, with the formation of the RHIC Spin Collaboration~\cite{PennState1990}, a group of experimental, theoretical and accelerator physicists with a common interest in spin, whose purpose was to add polarized proton capability to the Relativistic Heavy Ion Collider (RHIC). This idea came to fruition when, in 1995, RIKEN, the Institute of Physical and Chemical Research, in Japan decided to fund the spin hardware at RHIC and provide a second muon arm for the PHENIX experiment; and enhanced in 1997 when the RBRC (Fig.~\ref{fig:RSCoriginators}b) was founded at BNL, with T.~D.~Lee  as director, with research focus on spin physics, Lattice \QCD\  and Quark Gluon Plasma (\QGP) physics by nurturing a new generation of young physicists.   

The original goal of the RSC was that ``Operation of RHIC with two beams of highly polarized protons (70\%, either longitudinal or transverse) at high luminosity (${\cal L}=2\cdot 10^{32}$ cm$^{-2}$ sec$^{-1}$ for two months/year will allow high statistics studies of polatization phenomena in the perturbative region of hard scattering where both \QCD\ and EletroWeak theory make detailed prediction of polarization effects.'' It was expected that the integrated luminosity for two years (2$\times$2 months) at would be $\int {\cal L} dt=8\times 10^{38}$ cm$^{-2}$ (800 pb$^{-1}$) at $\sqrt{s}=500$ GeV~\cite{MJTICTP}. 
The principal physics results to be obtained were Spin Structure Functions which require measurements to complement DIS electron measurements: a) Gluon ($G(x)$) and Gluon spin ($\Delta G(x)$) structure functions by inclusive $\gamma$ and $\gamma$+Jet measurements; b) spin structure functions $\Delta{\bar{q}}$ from Drell-Yan, $\Delta{\bar{u}}$ from $W^{-}$ and $\Delta{\bar{d}}$ from $W^{+}$. 

Following Bourrely and Soffer~\cite{BS1993}, the single longitudinal spin parity violating asymmetry of the $W$ by flipping the spin of the proton is:
$A_L^W=(1/P)\times (\sigma^- -\sigma^+)/(\sigma^- + \sigma^+)$, where  $\sigma=d\sigma^{W}/dy$ and the $+$ or $-$ signs refer to the spin along or opposite to the direction of the proton with polarization $P$ (see Eq.~\ref{eq:ALW+}) where
$x_1={m_W\over \sqrt{s}}e^y$, $x_2={m_W\over \sqrt{s}}e^{-y}$.
   
%\vspace{0.2in}
\begin{center}
\begin{equation}
A_L^{W^+}(y)={ {-\Delta u(x_1,M_W^2) \bar{d}(x_2,M_W^2)+ 
\Delta \bar{d}(x_1,M_W^2) u(x_2,M_W^2)}
\over {u(x_1,M_W^2) \bar{d}(x_2,M_W^2)+ 
\bar{d}(x_1,M_W^2) u(x_2,M_W^2)} }
\label{eq:ALW+}
\end{equation}
\end{center}
and for $A_L^{W^-}(y)$ the $u\rightarrow d$ and $\bar{d}\rightarrow \bar{u}$.

\begin{figure}[!h]
\begin{center}
\raisebox{0.32pc}{{\footnotesize a)}\hspace*{-0.0pc}\includegraphics[width=0.59\textwidth]{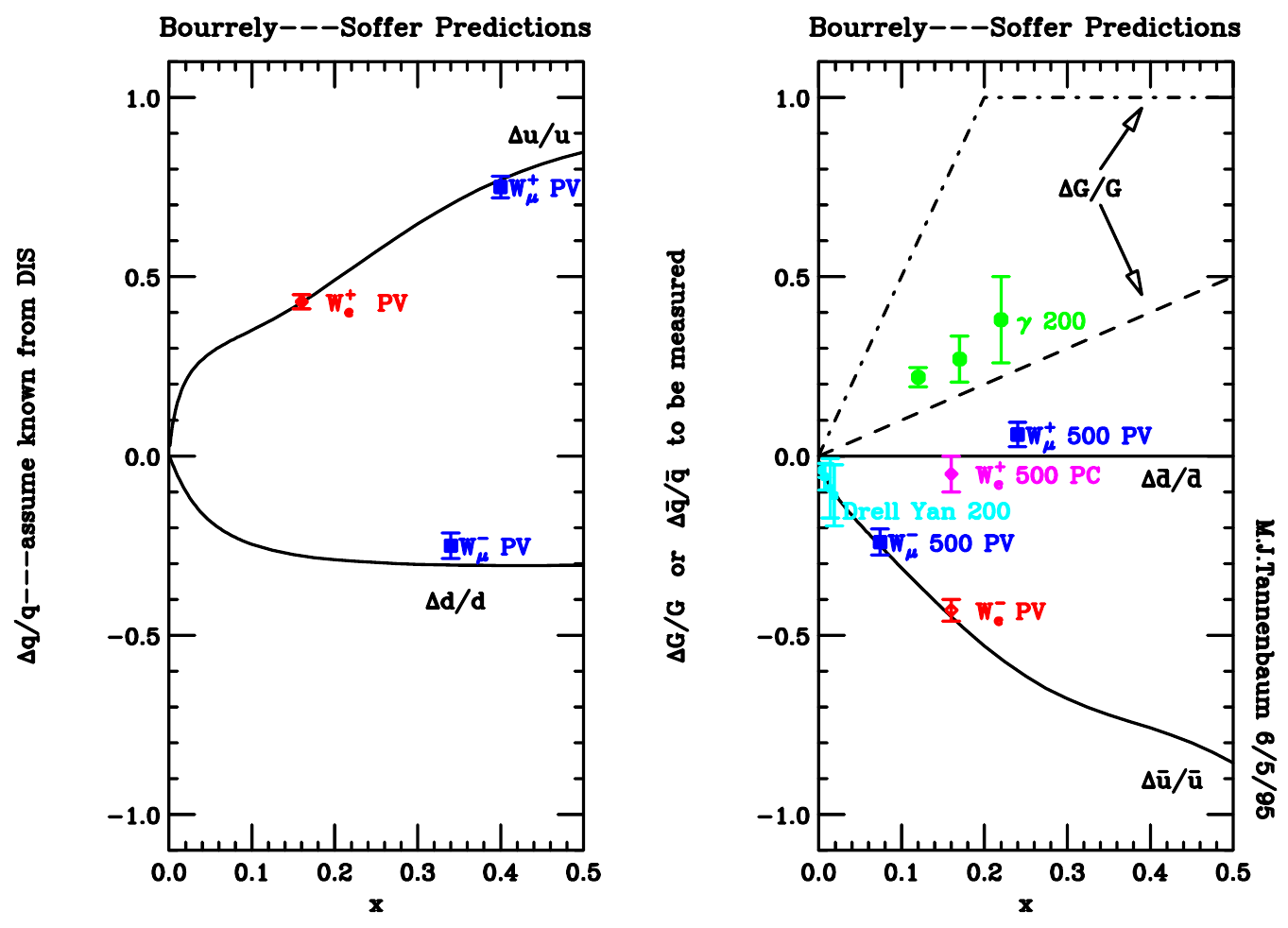}}\hspace*{0.2pc}
\raisebox{0.0pc}{{\footnotesize b) }\hspace*{-0.1pc}\includegraphics[width=0.37\textwidth]{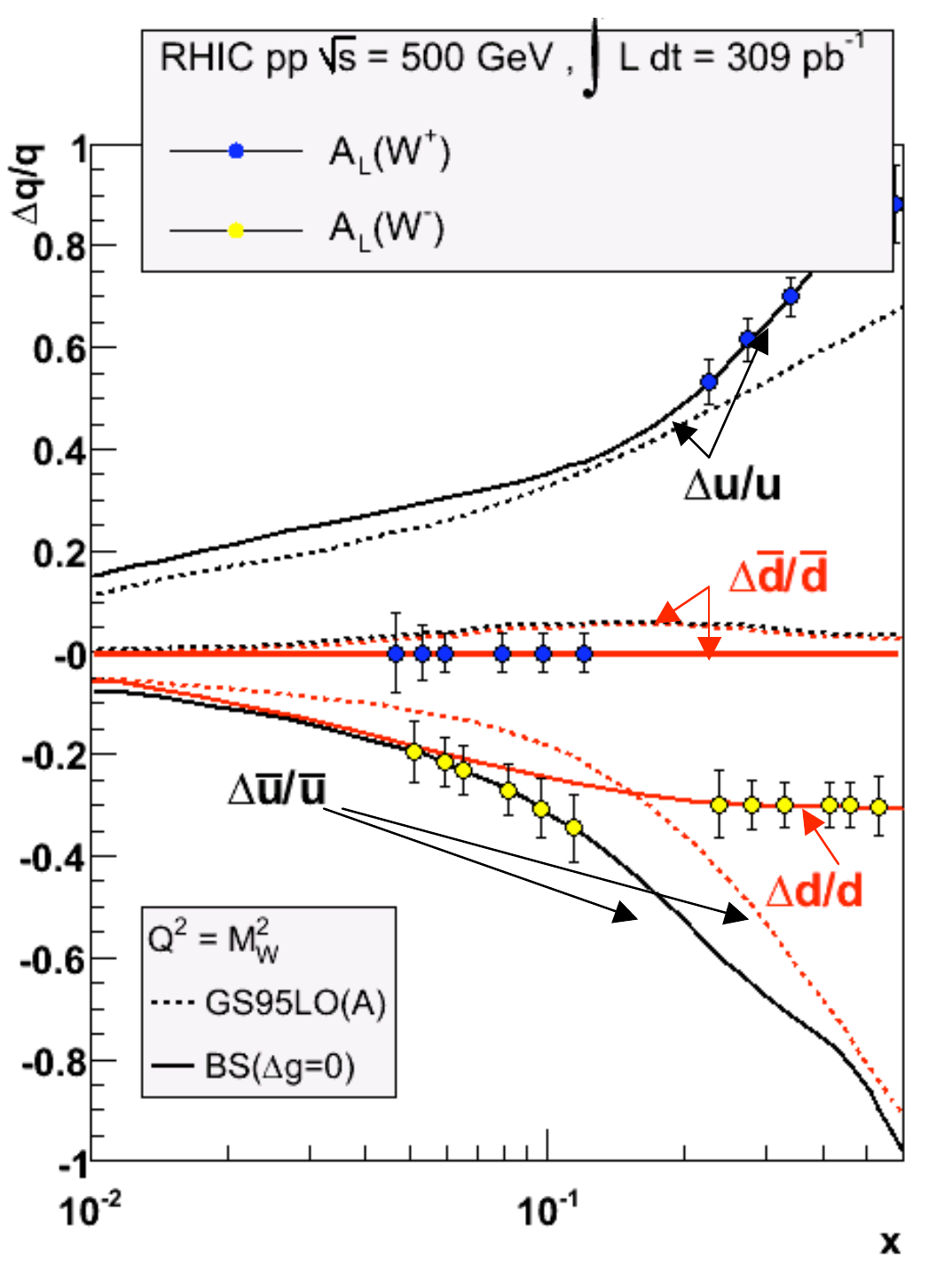}}
\end{center}\vspace*{-1.5pc}
\caption[]{\footnotesize Expected results from $W^{\pm}$ parity violating spin asymmetry, anti-quark and Gluon spin asymmetry c.1995: a) favoring mid-rapidity~\cite{MJTICTP} with 800 pb$^{-1}$ integrated luminosity at $\sqrt{s}=500$ GeV; b) favoring forward rapidity (only 309 pb$^{-1}$)~\cite{NaohitoICTP}. Signs are reversed for $W^{\pm}$ compared to Eq.~\ref{eq:ALW+}. }
\label{fig:RSCexpected}\vspace*{-0.5pc}
\end{figure}

The plan in PHENIX was to detect the reaction $u+\bar{d}\rightarrow W^+ \rightarrow e^+ + \nu_e$ at mid-rapidity and the decay $W^+ \rightarrow \mu^+ +\nu_{\mu}$ at forward rapidity $1.1<|y|<2.3$ and we thought that we could calculate the $x$ of the $W$ in these reactions. This led to some nice predictions for the results, circa 1995 (Fig.~\ref{fig:RSCexpected}). 

\begin{figure}[!h]
\begin{center}
\raisebox{1.0pc}{\includegraphics[width=0.50\textwidth]{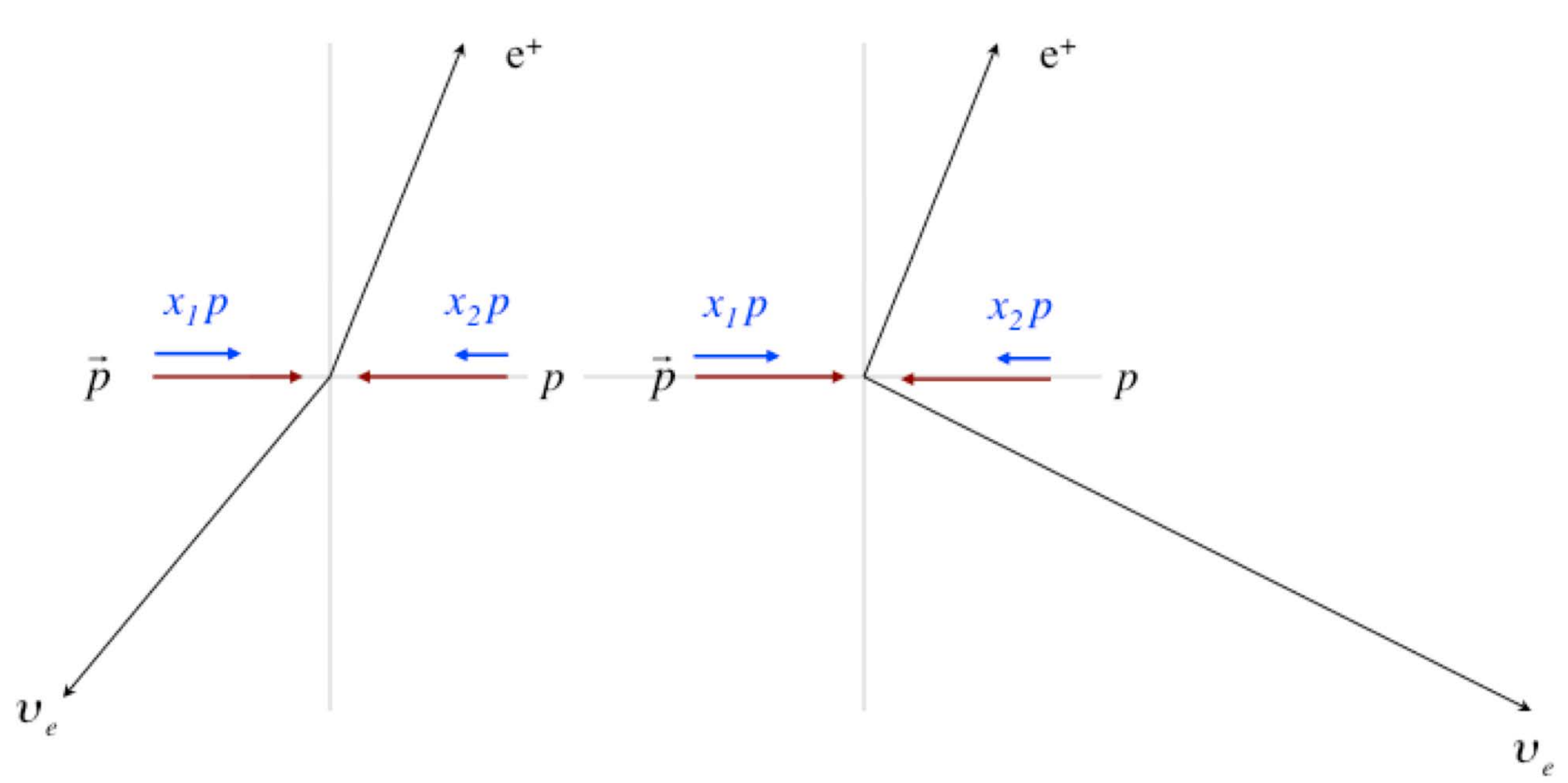}}\hspace*{0.2pc}
\raisebox{0pc}{\includegraphics[width=0.50\textwidth]{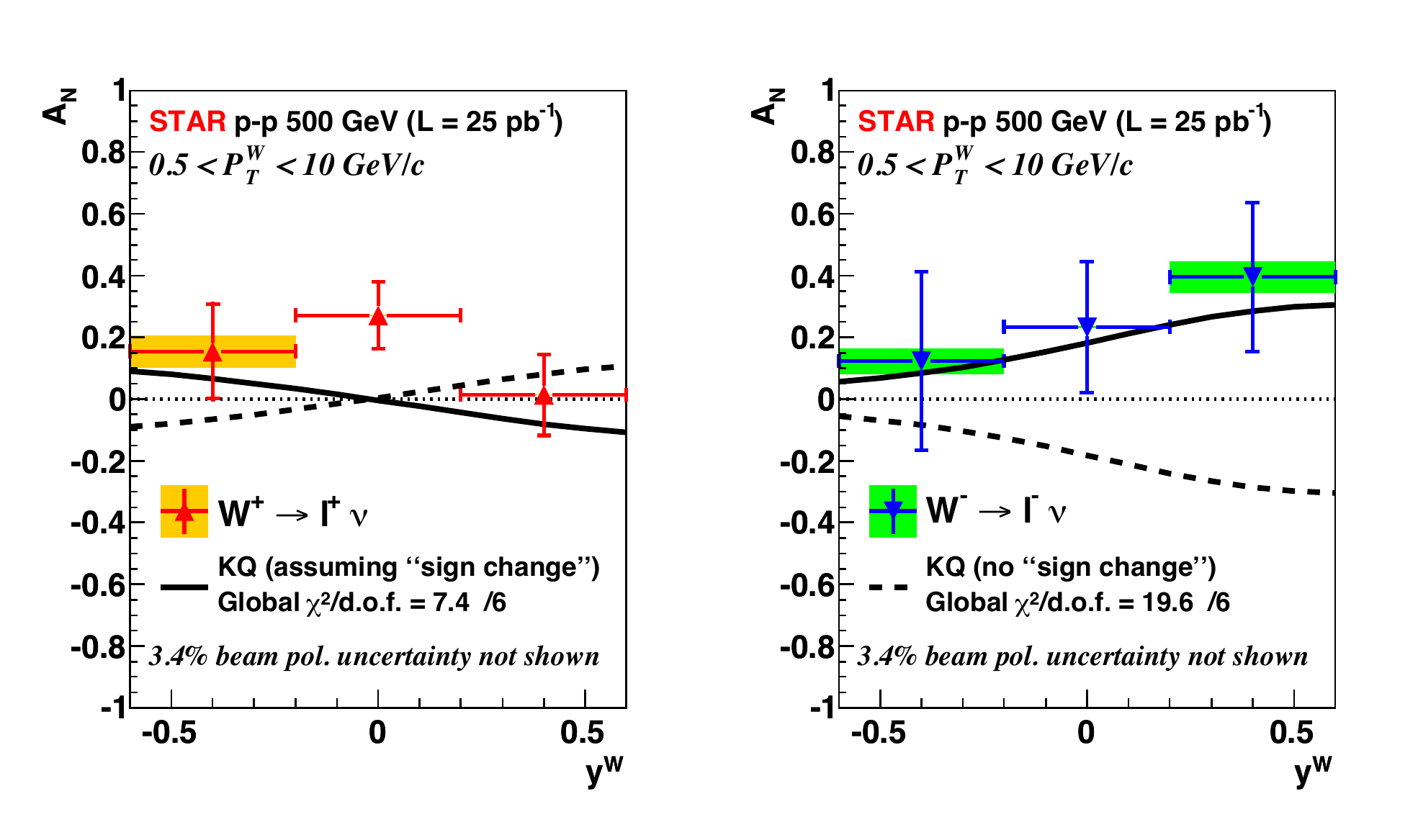}}
\end{center}\vspace*{-1.5pc}
\caption[]{\footnotesize a) (left) Kinematic ambiguity of $y^W$ for $W^{+}\rightarrow e^{+} + \nu_e$ with $y_e$=+0.35, $p_T^e=35$ GeV/c: either $y_{\nu_e}=-0.73$ ($y^W=0.19$) or $y_{\nu_e}=1.43$ ($y^W=+0.89$). b) (right) Transverse single spin asymmetry as a function of rapidity $y^W$ for $W^{+}$ and $W^{-}$ from Run-11 $\int{\cal L} dt=25$ pb$^{-1}$~\cite{STARPRL116}.  }
\label{fig:ANW}\vspace*{-0.5pc}
\end{figure}

We thought that we could calculate $x_1$ and $x_2$ in p$+$p$\rightarrow W^{\pm}+X$, followed by the leptonic decay. This works reasonably well for the $\mu$ at forward rapidity but there is a kinematic ambiguity for smaller rapidities (Fig.~\ref{fig:ANW}) as to whether the $W$ is in the same or opposite direction to the $e^{\pm}$. 
However, this posed no problem for either PHENIX~\cite{PXW} or STAR~\cite{STW} to measure $W^{\pm}$ production for $e^{\pm}$ at mid-rapidity with the Zichichi Jacobian peak~\cite{AZ1964}. 

Last year, STAR seems to have overcome the ambiguity with a measurement of the single transverse spin asymmetry $A_N$ of the $W^+$ and $W^{-}$ as a function of $y_W$ over the range $|y_W|\leq 0.6$ (Fig.~\ref{fig:ANW}b)~\cite{STARPRL116}. This measurement tests Transverse Momentum Dependent (TMD) parton distribution functions with respect to the `intrinsic' transverse momentum $k_T$ of partons in the proton. For example, the Sivers function is the correlation of $k_T$ of a parton with the spin of the proton which may change sign according to the number of gluons exchanged in the reaction. 
Clearly this measurement from Run-11 is inadequate to draw a firm conclusion; so the answer to the question following Fig.~\ref{fig:RHICruns} (why transverse polarization to study TMD effects in Run-17 when the main thrust of the RSC is longitudinal polarization) is to get more data for Fig.~\ref{fig:ANW}b, with $\approx 100$ pb$^{-1}$ collected. 

For longitudinally polarized protons, the total collected luminosity at RHIC in Fig.~\ref{fig:RHICruns} is $\approx 200$ pb$^{-1}$, so decent results as in Fig.~\ref{fig:RSCexpected} should be forthcoming. However there may be a problem for future runs: the new U.S. President's proposed budget for 2018 terminates the RHIC spin program in order to fund the `higher priority' 12 GeV JLAB science program. The budget also proposes to end U.S. participation in the LHC heavy ion program to focus funding on RHIC. The good news is that the U.S. congress has not yet acted on this budget at the present writing (November 2017) so these activities will continue until further notice (and hopefully beyond).   

\section{Heavy Ion Physics results at RHIC this past year.}
Since the startup of RHIC in the year 2000, many discoveries have been made including the \QGP\ as the perfect liquid. I present a quick summary and then move on to the latest results.
\begin{itemize}
\item  Suppression of high $p_T$ hadrons from hard-scattering of initial
state partons; also modification of the away-side jet. \vspace*{-0.5pc}
\item Elliptic Flow at the Hydrodynamic limit with shear viscosity/entropy density at or near the quantum lower bound $\eta/s=1/(4\pi)$ $\Longrightarrow$ \QGP\ the Perfect Liquid. \vspace*{-0.5pc}
\item Elliptic flow of particles proportional to the number of the valence (constituent) quark count.\vspace*{-0.5pc}
\item Charged particle multiplicity proportional to the number of constituent quark participants.\vspace*{-0.5pc}
\item Higher order flow moments proportional to density fluctuations of the initial colliding nuclei.\vspace*{-0.5pc}
\item Suppression and flow of heavy quarks roughly the same as that of light quarks; \QCD\ hard direct photons not suppressed, don{'}t flow.\vspace*{-0.5pc}
\item Production and flow of soft photons $p_T<2$ GeV/c with exponential distribution.\vspace*{-0.5pc}
\end{itemize}
\subsection{Flow in small systems}
\begin{figure}[!h]
\begin{center}
\raisebox{1.0pc}{{\footnotesize a)}\hspace*{-0.0pc}\includegraphics[width=0.44\textwidth]{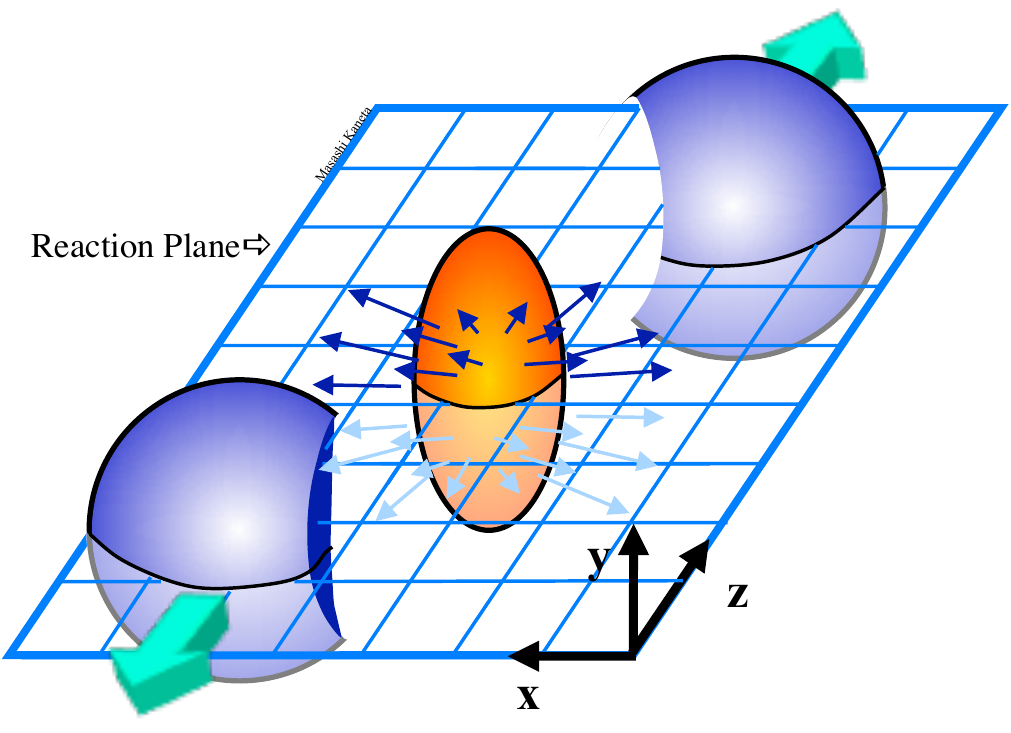}}\hspace*{0.2pc}
\raisebox{0pc}{{\footnotesize b)}\hspace*{-0.0pc}\includegraphics[width=0.51\textwidth]{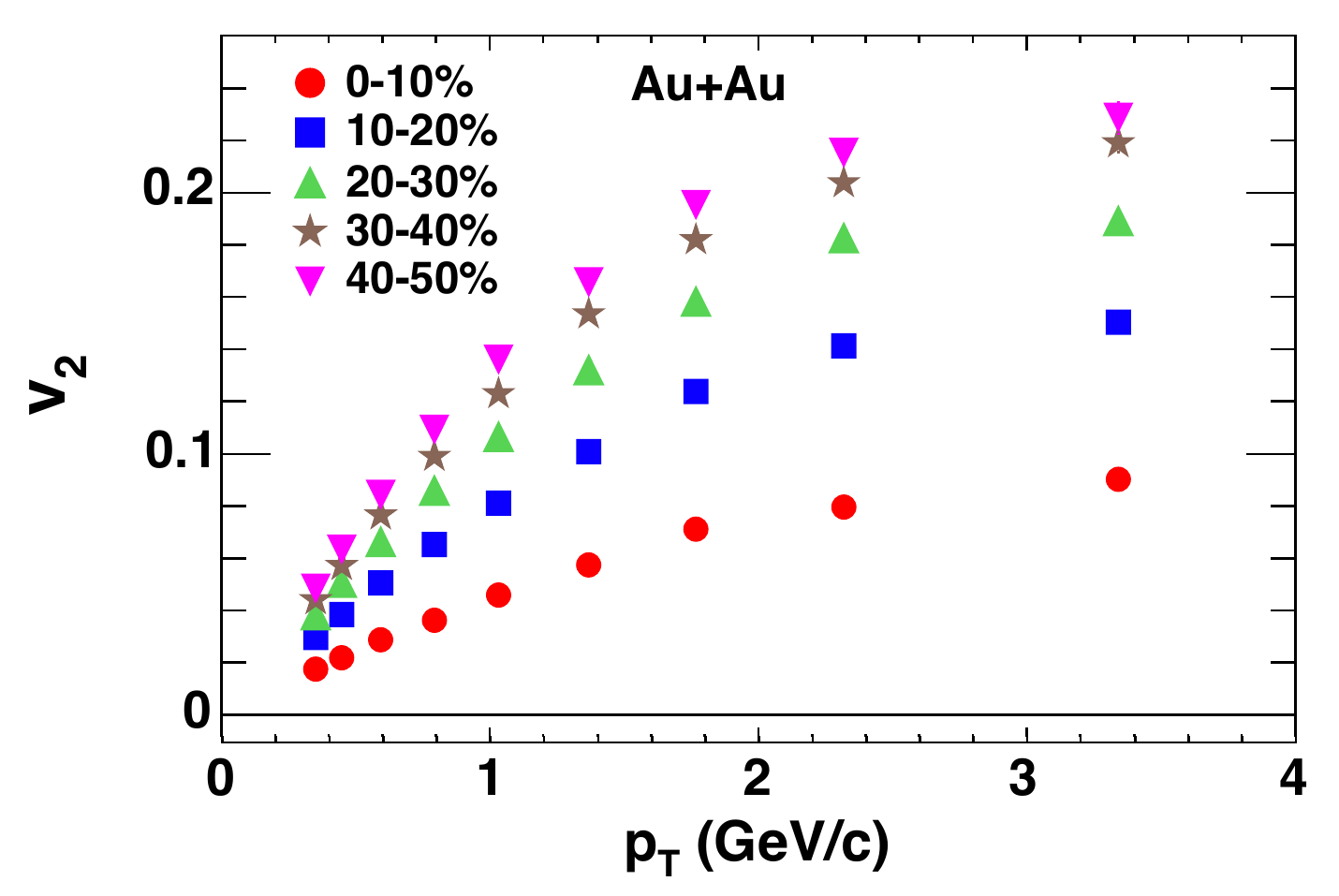}}
\end{center}\vspace*{-1.5pc}
\caption[]{\footnotesize a) Almond shaped overlap zone generated just after an A$+$A collision where the incident nuclei are moving along the $\pm z$ axis. The reaction plane by definition contains the impact parameter vector (along the $x$ axis, which defines $\phi=0$)~\cite{KanetaQM04}. b) $v_2$ as a function of $p_T$ for the centralities (0-10\% is most central) indicated in Au$+$Au collisions at $\sqsn=200$ GeV~\cite{ppg062}.}
\label{fig:AuAuFlow}\vspace*{-0.5pc}
\end{figure}

It was thought that elliptical flow, ($v_2=\mean{\cos2\phi}$), the emission of particles preferably along the short axis from the almond shape overlap region of the struck nucleons in an A$+$A collision (Fig.~\ref{fig:AuAuFlow}a) was an indication of a collective effect with hydrodynamic behavior related to the perfect liquid \QGP. The results from Au$+$Au collisions at $\sqsn=200$ GeV as a function of centrality (upper percentile) (Fig.~\ref{fig:AuAuFlow}b)~\cite{ppg062} clearly indicate that $v_2$ decreases with increasing centrality of the collision as the overlap region becomes less elliptical. 

\begin{figure}[!h]
\begin{center}
\raisebox{0pc}{\includegraphics[width=0.77\textwidth]{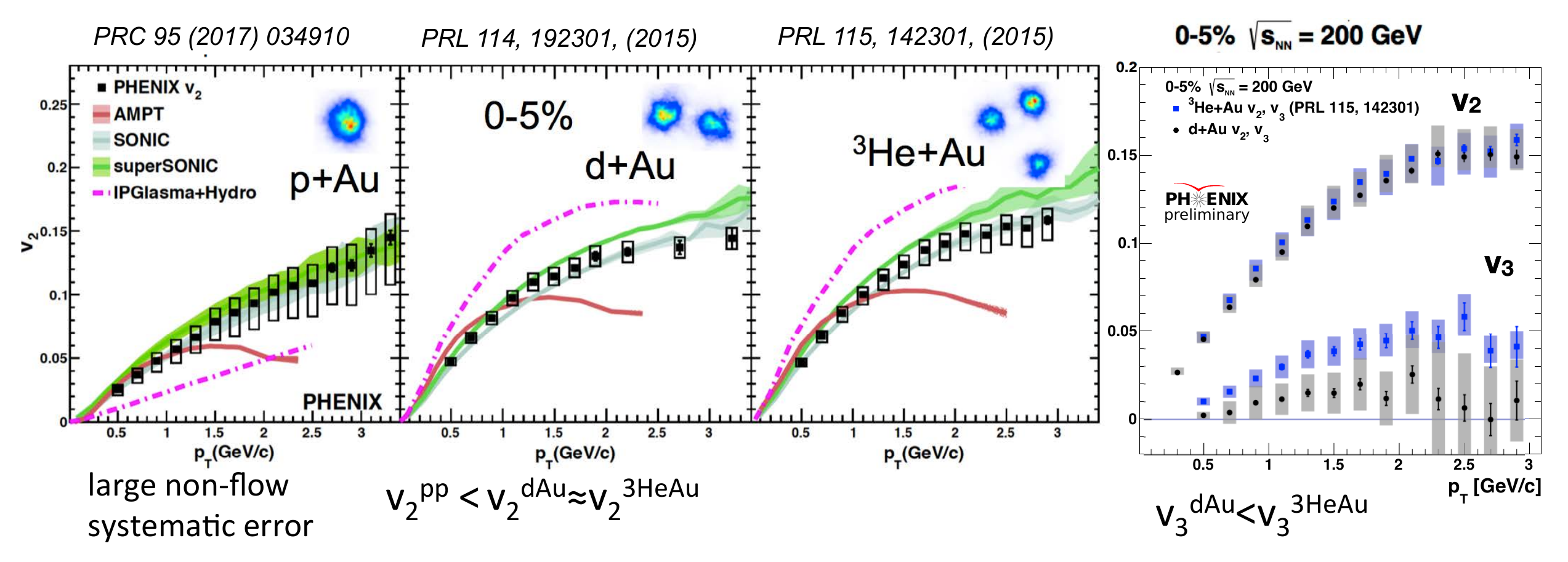}}
%\raisebox{0pc}{\includegraphics[width=0.77\textwidth]{figs/v2dau}}
\end{center}\vspace*{-1.5pc}
\caption[]{\footnotesize Anisotropic transverse flow the in small systems indicated.}
\label{fig:LightFlow}\vspace*{-0.5pc}
\end{figure}

Measurements of small systems at RHIC namely d$+$Au and p$+$Au were initially used to establish baseline nuclear effects in which neither hot nuclear matter nor the \QGP\ are produced~\cite{PXPRL91dAu}. However because $v_2$ depends on the geometry of the overlap region, it was decided to study whether small systems, p$+$Au, d$+$Au and $^3$He$+$Au, produce collective expansion and flow related to their different collision geometries~\cite{JamieNPRL113}. Recent results by PHENIX (Fig.~\ref{fig:LightFlow}) show that p$+$Au, d$+$Au and $^3$He$+$Au with distinctly different initial geometries have similar if not identical values of $v_2$ as a function of $p_T$ in collisions at $\sqsn=200$ GeV and centrality 0--5\%. Furthermore, $^3$He$+$Au has a significant triangular flow, $v_3=\mean{\cos3\phi}$, which is significantly larger than the preliminary measurement of $v_3$ in d$+$Au. 

These results provide clear evidence that the $v_2$ measured in small systems arises from initial geometry coupled to interactions between medium constituents resulting in collective expansion.  
Especially noteworthy is that the $v_2$ vs. $p_T$ values in the small systems at 0-5\% centrality (Fig.~\ref{fig:LightFlow}) are actually greater than the $v_2$ vs. $p_T$ values in Au$+$Au at comparable centrality (Fig.~\ref{fig:AuAuFlow}b), with obvious implication that there is still much to be learned about what is called `flow' in collisions with nuclei.

\subsection{A new detector and nice $v_2$ measurements of open charm (D$^0$) in Au$+$Au collisions by STAR, but \ldots}

\begin{figure}[!h]
\begin{center}
\raisebox{0.0pc}{{\footnotesize a)}\hspace*{-0.0pc}\includegraphics[width=0.49\textwidth]{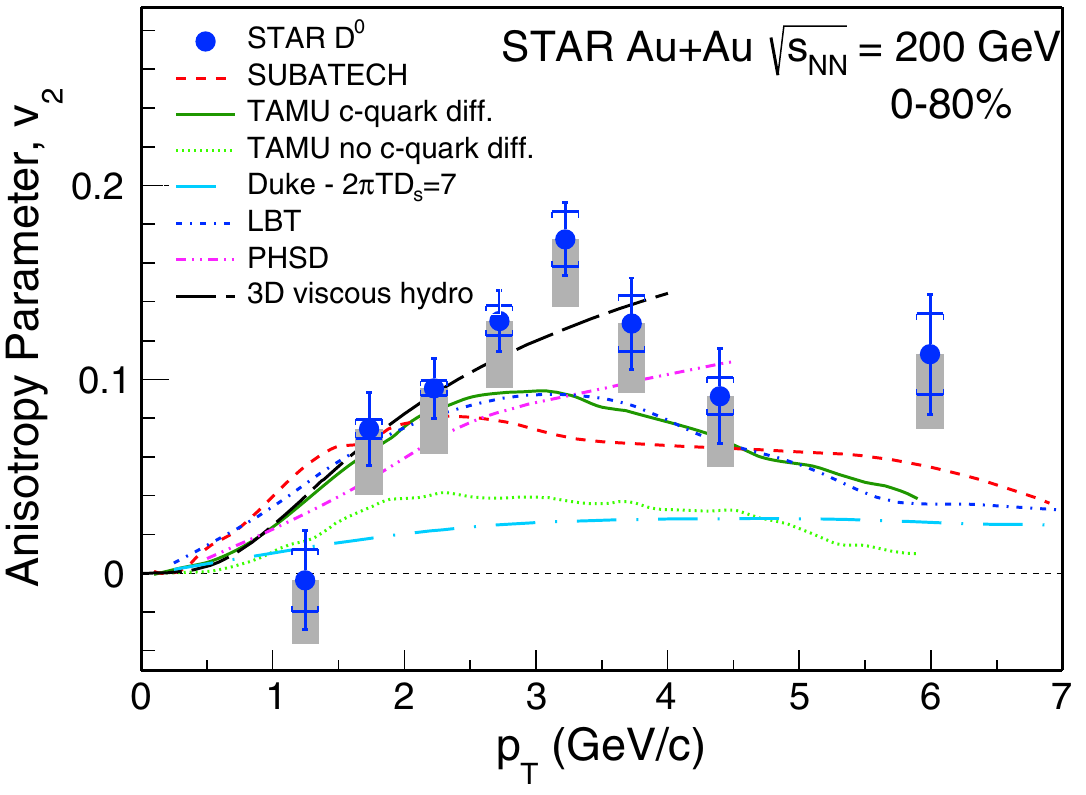}}\hspace*{0.2pc}
\raisebox{2.0pc}{{\footnotesize b)}\hspace*{-0.0pc}\includegraphics[width=0.44\textwidth]{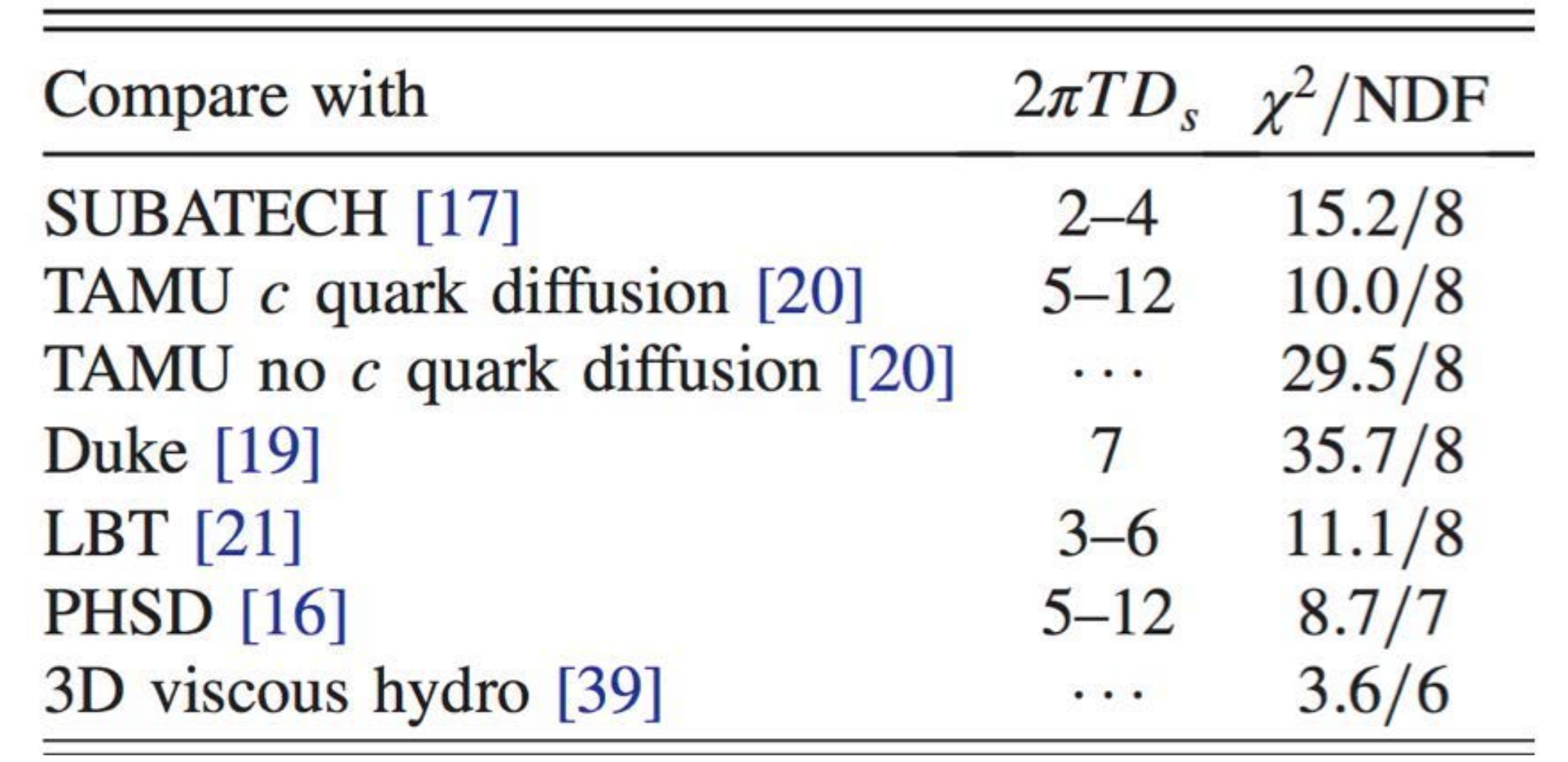}}
\end{center}\vspace*{-1.5pc}
\caption[]{\footnotesize a) STAR $v_2$ of D$^0$ meson with theory calculations~\cite{STARDPRL118}. b) Table of Diffusion coefficients of theory calculations shown in (a) referring to numbered references in~\cite{STARDPRL118}.}\label{fig:STARDflow}\vspace*{-0.5pc}
\end{figure}
The first results from the STAR Heavy Flavor Tracker (HFT), a state of the art silicon vertex detector which includes the first use of Monolithic Active Pixel Sensors (MAPS), were published this year, a measurement of $v_2$ in open charm $D^0$ mesons~\cite{STARDPRL118} (Fig.~\ref{fig:STARDflow}a). Their conclusion was that:``Several theoretical calculations with temperature-dependent, dimensionless charm quark spatial diffusion coefficients ($2\pi T D_s$) in the range of $\sim2-12$ (Fig.~\ref{fig:STARDflow}b) can simultaneously reproduce our D$^0$ $v_2$ result as well as the previously published STAR measurement of the D$^0$ nuclear modification factor~\cite{STARDPRL113}".  
\subsubsection{\ldots but PHENIX did this 10 years ago with prompt single $e^{\pm}$ from charm with numerical results: $\eta/s=(\frac{4}{3} \ {\rm to\ }2)/4\pi$ $\Longrightarrow$ \QGP\ the Perfect Liquid.}
\begin{figure}[!h]  
\begin{center}
\raisebox{+0.0pc}{\includegraphics[width=0.48\textwidth]{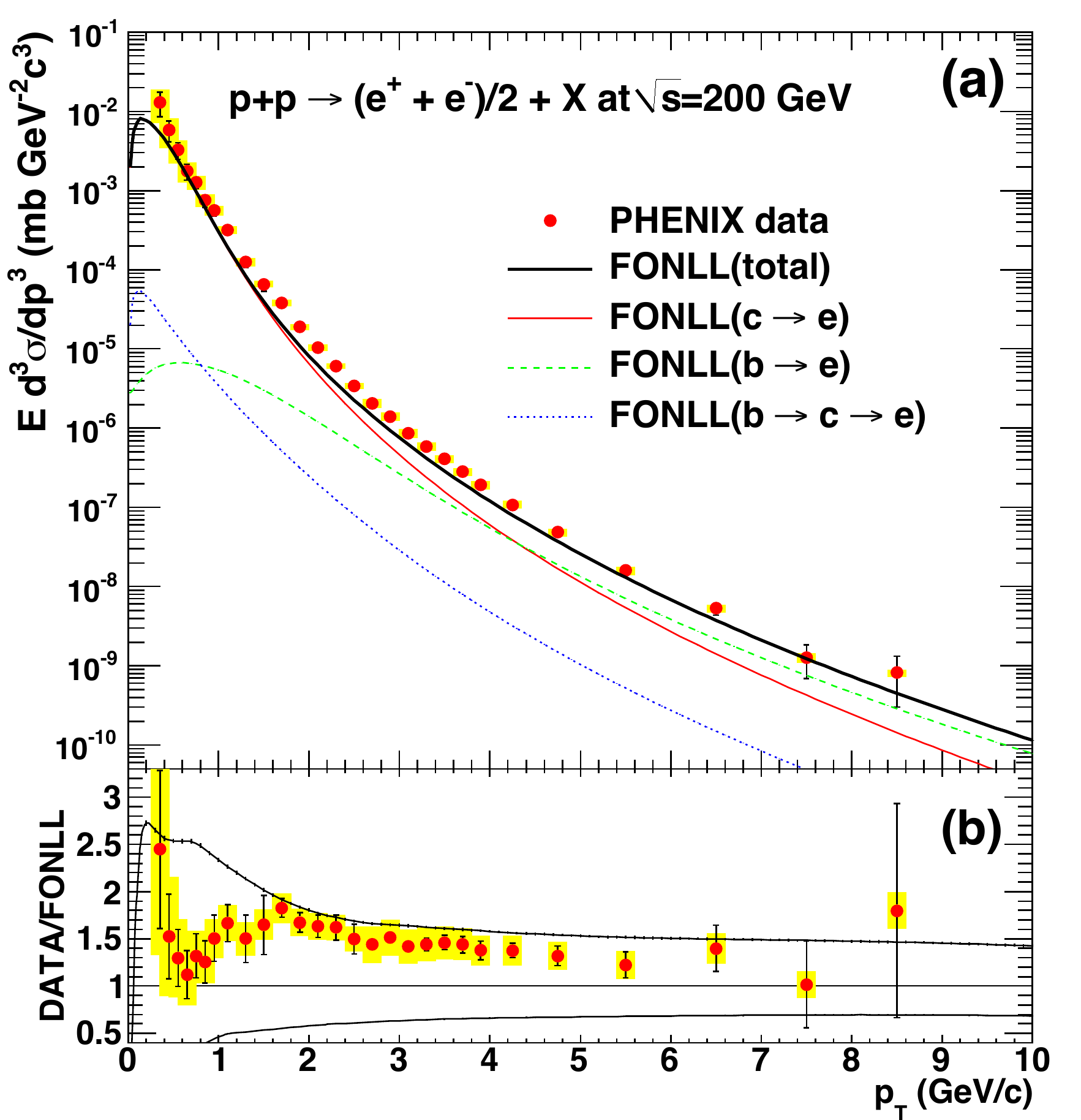}}
\raisebox{-0.2pc}{\includegraphics[width=0.51\textwidth]{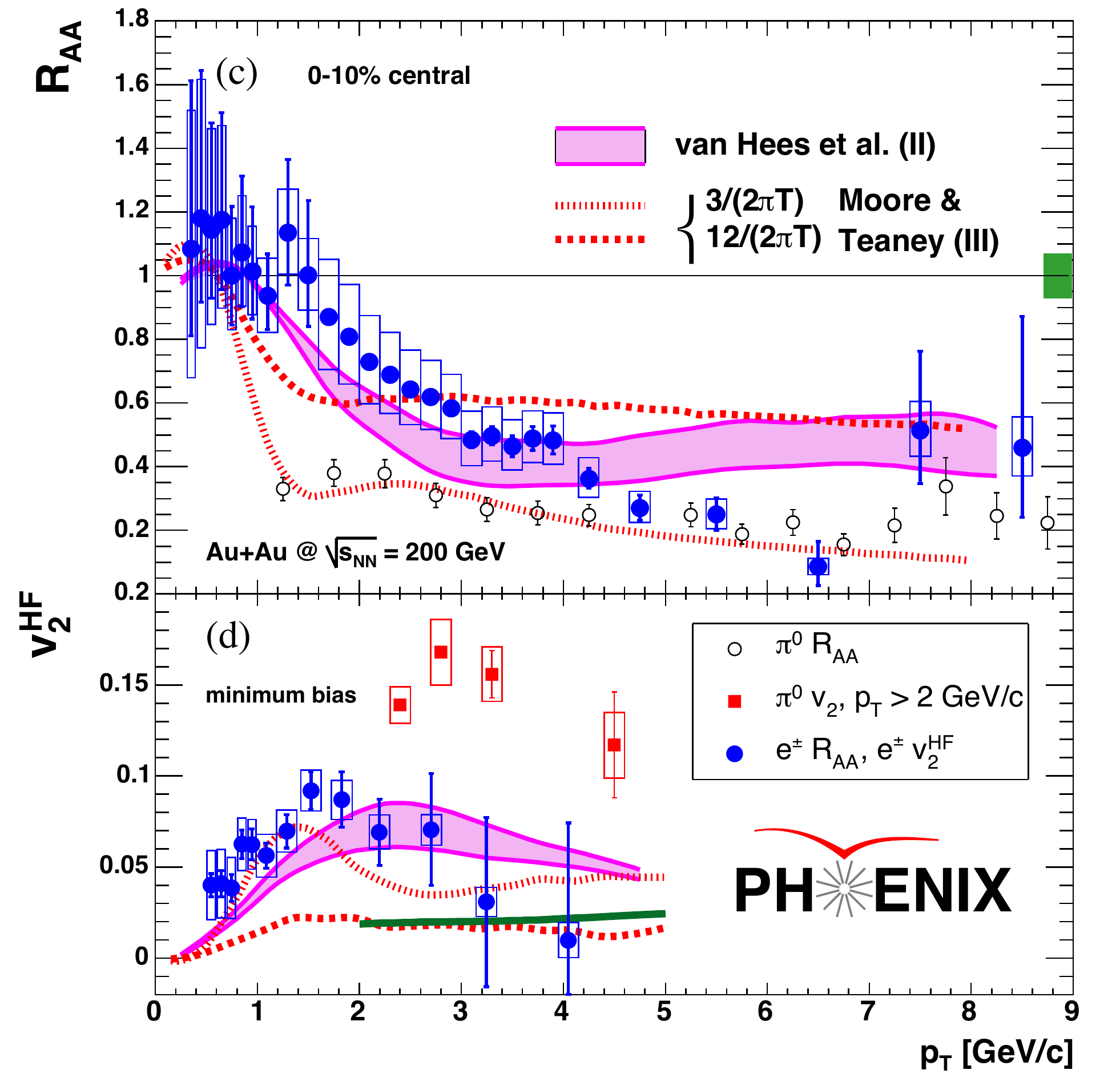}}
\caption[]{\footnotesize a) Invariant differential cross sections of electrons from heavy flavor quark decays~\cite{ppg065}. The curves are FONNL calculations~\cite{RamonaFONNL2006} b) The ratio of the measurement to the FONNL calculation. c) $R_{AA}$ from direct-single $e^{\pm}$ and $\pi^0$~\cite{ppg066} and d) elliptical flow $v_2$ for Heavy Flavor quarks and $\pi^0$. Curves are theoretical predictions with the diffusion coefficient D~\cite{MandTPRC71},\cite{vanHees}.}   
\label{fig:PerfectLiquid}%\vspace*{-0.12in}
\end{center}
\end{figure}

In Fig.~\ref{fig:PerfectLiquid}a, the PHENIX measurement~\cite{ppg065} of the $p_T$ spectrum of prompt single $e^{\pm}$ from charm decay at mid-rapidity in p$+$p collisions at $\sqrt{s}=200$ GeV  is presented along with theory calculations in fixed-order-plus-next-to-leading-log (FONNL) p\QCD~\cite{RamonaFONNL2006}, with the contributions from $b$ and $c$ quark decay indicated, in excellent agreement with the measurement. Fig.~\ref{fig:PerfectLiquid}c~\cite{ppg066} shows the suppression of single $e^{\pm}$ in 0-10\% centrality Au$+$Au collisions which is quantified by the nuclear modification factor $R_{AA}\equiv dN_{A+A}/(\mean{T_{AA}}d\sigma_{p+p})$, where $dN_{A+A}$ is the differential yield in A$+$A collisions, $d\sigma_{p+p}$ is the differential cross section in p$+$p collisions at the same $p_T$, and $\mean{T_{AA}}$ is the average overlap integral of the nuclear thickness functions for the given centrality. Fig.~\ref{fig:PerfectLiquid}d shows the measured $v_2^{\rm HF}$ for the single $e^{\pm}$ from Heavy Flavor. Both $R_{AA}$ and $v_2^{\rm HF}$ are compared to that of $\pi^0$.   

The observed flow of heavy quarks led Moore and Teaney~\cite{MandTPRC71} to suggest that the medium responds as a thermalized fluid and that the transport mean free path is small. They treated the heavy quark thermalizing as a diffusion problem with diffusion coefficient $D\approx6\eta/(\varepsilon +p)$. The enthalpy $\varepsilon+p$ equals ${Ts}$ at baryon chemical potential $\mu_B=0$ so that $D=6\eta/(Ts)$. When combined with the values $D=(6\ {\rm to}\ 4)/(2{\pi}T)$ from curve II~\cite{vanHees} on Fig.~\ref{fig:PerfectLiquid}c,d the result is a value of $\eta/s\approx (2\ {\rm to}\ 4/3)/(4\pi)$, intriguingly close to the conjectured quantum lower bound~\cite{KSSPRL94} $\eta/s\approx 1/(4\pi)$, hence the perfect liquid.

\subsection{Jet Quenching: the first \QCD\ based prediction BDMPSZ~\cite{BSZ2000}} 
\begin{figure}[!h]  
\begin{center}
\raisebox{+2.0pc}{{\footnotesize a)}\hspace*{-0.1pc}\includegraphics[width=0.42\textwidth]{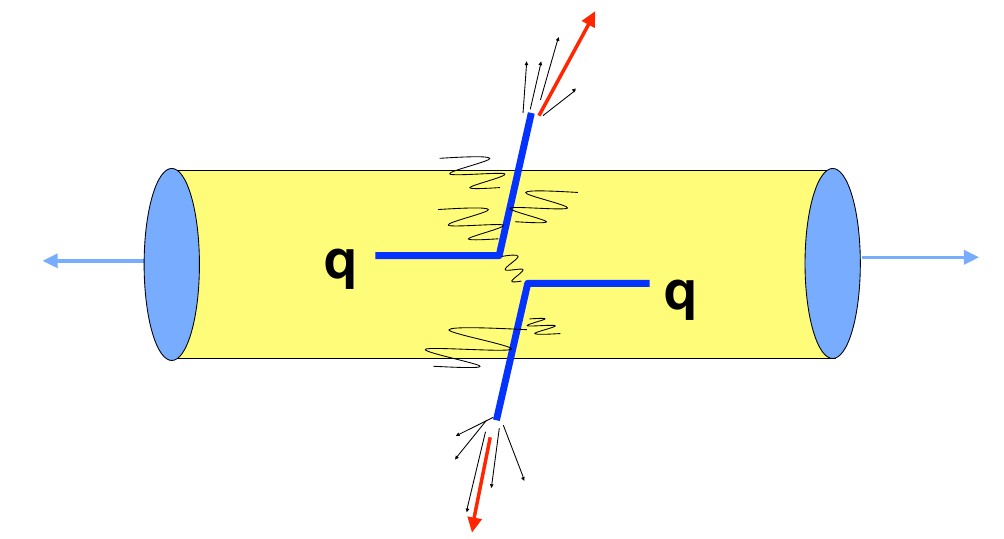}}
\raisebox{-0.0pc}{{\footnotesize b)}\hspace*{-0.2pc}\includegraphics[width=0.54\textwidth]{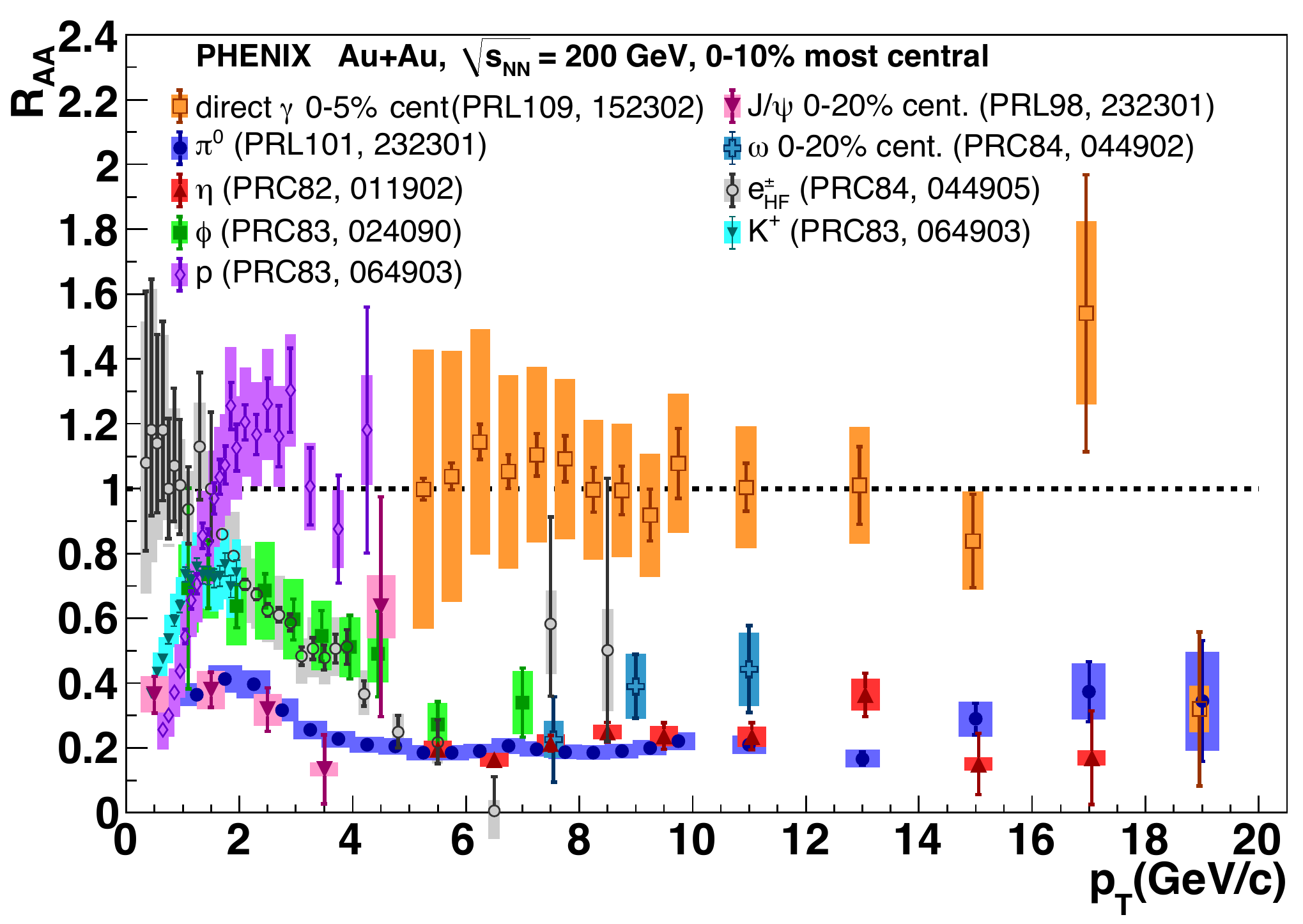}}
\caption[]{\footnotesize a) Schematic of $q+q$ scattering with scattered quarks losing energy in the medium. b) $R_{AA}(p_T)$ for all identified particles so far measured by PHENIX in Au$+$Au central collisions at $\sqsn=200$ GeV.}   
\label{fig:suppression}%\vspace*{-0.12in}
\end{center}
\end{figure}

The first prediction of how to detect the \QGP\ was via $J/\Psi$ suppression~\cite{MatsuiSatz} in 1986 (see Fig.~\ref{fig:70-2}). However the first \QCD\ based prediction for detecting the \QGP\ was BDMPSZ Jet Quenching~\cite{BSZ2000}.  This is produced by the energy loss, via LPM coherent radiation of gluons, of an outgoing parton with color charge fully exposed in a medium with a large density of similarly exposed color charges (i.e. the \QGP) (Fig.~\ref{fig:suppression}a). Jet quenching was observed quite early at RHIC by suppression of high $p_T$ $\pi^0$~\cite{PXPRL88},  with lots of subsequent evidence (Fig.~\ref{fig:suppression}b). It is interesting to note that all identified hadrons generally have different $R_{AA}$ for $p_T\leq5$ GeV/c but tend to converge to the same value for $p_T\gsim5$ GeV/c. The fact that direct-$\gamma$ are not suppressed indicates that suppression is a medium effect on outgoing color-charged partons as predicted by BDMPSZ~\cite{BSZ2000}.  
\subsubsection{But the BDMPSZ model has two predictions}
\begin{figure}[!h]  
\begin{center}
\raisebox{+0.0pc}{\includegraphics[width=0.60\textwidth]{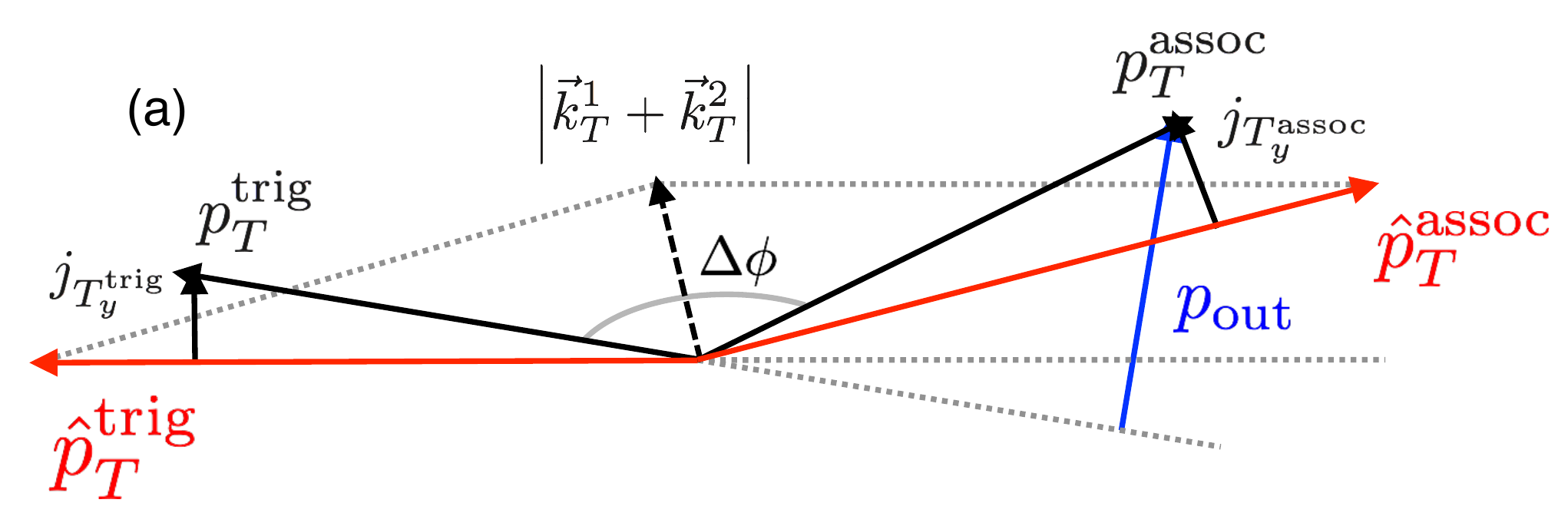}}\hspace*{0.2pc}
\caption[]{\footnotesize a) Initial configuration: trigger jet $\hat{p}_{Tt}$ associated (away) jet $\hat{p}_{Ta}$  with $k_T$ effect (dashed arrow) and fragments $p_{Tt}$ and $p_{Ta}$, with fragmentation transverse momentum $j_{T_y}$, and $p_{out}=p_{Ta}\sin(\pi-\Delta\phi)$.   }   
\label{fig:kTprime}
\end{center}
\end{figure}

\noindent(I) The energy loss of the outgoing parton, $-dE/dx$,  
per unit length ($x$) of a medium with total length $L$, is proportional to the total 4-momentum transfer-squared, $q^2(L)$, with the form:\vspace*{-0.5pc}
\begin{equation}{-dE \over dx }\simeq \alpha_s \langle{q^2(L)}\rangle=\alpha_s\, \mu^2\, L/\lambda_{\rm mfp} 
=\alpha_s\, \hat{q}\, L\qquad \label{eq:dEdx} \end{equation}
where $\mu$, is the mean momentum transfer per collision, and the transport coefficient 
{$\hat{q}=\mu^2/\lambda_{\rm mfp}$} is the 4-momentum-transfer-squared to the medium per mean free path, $\lambda_{\rm mfp}$.\\[0.5pc] 
\noindent(II) Additionally, the accumulated momentum-squared, $\mean{p^2_{\perp W}}$ transverse to a parton traversing a length $L$ in the medium  is well approximated by\vspace*{-0.5pc} 
\begin{equation}\mean{p^2_{\perp W}}\approx\langle{q^2(L)}\rangle=\hat{q}\, L \qquad \mbox{\rm so that}\qquad  \mean{\hat{q} L}/2=\mean{k_{T}^2}_{AA}-\mean{k{'}_{T}^2}_{pp} \label{eq:broadening}\end{equation} since only the component of $\mean{p^2_{\perp W}}$ $\perp$ to the scattering plane affects $k_T$. This is called azimuthal broadening. Here (see Fig.~\ref{fig:kTprime}) $k_T$ denotes the intrinsic transverse momentum of a parton in a proton plus any medium effect and $k{'}_T$ denotes the reduced value correcting for the lost energy of the scattered partons in the \QGP, a new idea this year~\cite{MJTPLB771}.
%\vspace*{-2.0pc} 

Even though jet quenching has been established and confirmed for more than 15 years, many experiments have tried to find azimuthal broadening at RHIC e.g. \cite{STARJhPRL112}, \cite{JacobsNPA956}, but have not been able to observe the effect because of systematic uncertainties.
\subsubsection{ Understanding $k_T$ and $k{'}_T$.}
In Fig.~\ref{fig:kTprime}a, following the methods of Feynman,Field and Fox~\cite{FFFNPB128}, CCOR~\cite{CCORjTkT} and PHENIX~\cite{ppg029}, the $\mean{k^2_T}$ for di-hadrons is computed from Fig.~\ref{fig:kTprime}a as:
\begin{equation}\sqrt{\mean{k^2_T}}=\frac{\hat{x}_h}{\mean{z_t}}\sqrt{\frac{ \mean{p^2_{\rm out}}-(1+{x_h^2})\rms{j_T}/2}{x_h^2}} \label{eq:kTcalc}\end{equation}
where ${p}_{Tt}$, ${p}_{Ta}$ are the transverse momenta of the trigger and away particles, \mbox{$x_h=p_{Ta}/p_{Tt}$}, $\Delta\phi$ is the angle between ${p}_{Tt}$ and ${p}_{Ta}$ and  $p_{\rm out}\equiv p_{Ta} \sin(\pi-\Delta\phi)$.  The di-hadrons are assumed to be fragments of jets with transverse momenta $\hat{p}_{Tt}$ and $\hat{p}_{Ta}$ with ratio $\hat{x}_h=\hat{p}_{Ta}/\hat{p}_{Tt}$. ${z_t}\simeq p_{Tt}/\hat{p}_{Tt}$ is the fragmentation variable, the fraction of momentum of the trigger particle in the trigger jet. $j_T$ is the jet fragmentation transverse momentum and we have taken $\rms{\jt{ay}}\equiv\rms{\jt{a\phi}}=\rms{\jt{t\phi}}=\rms{j_T}/{2}$. The variable $x_h$ (which STAR calls $z_T$) is used as an approximation of the variable $x_E=x_h\cos(\pi-\Delta\phi)$ from the original terminology at the CERN ISR where $k_T$ was discovered and measured 40 years ago.

\begin{figure}[!b]  
\begin{center}
\raisebox{+0.0pc}{\includegraphics[width=0.88\textwidth]{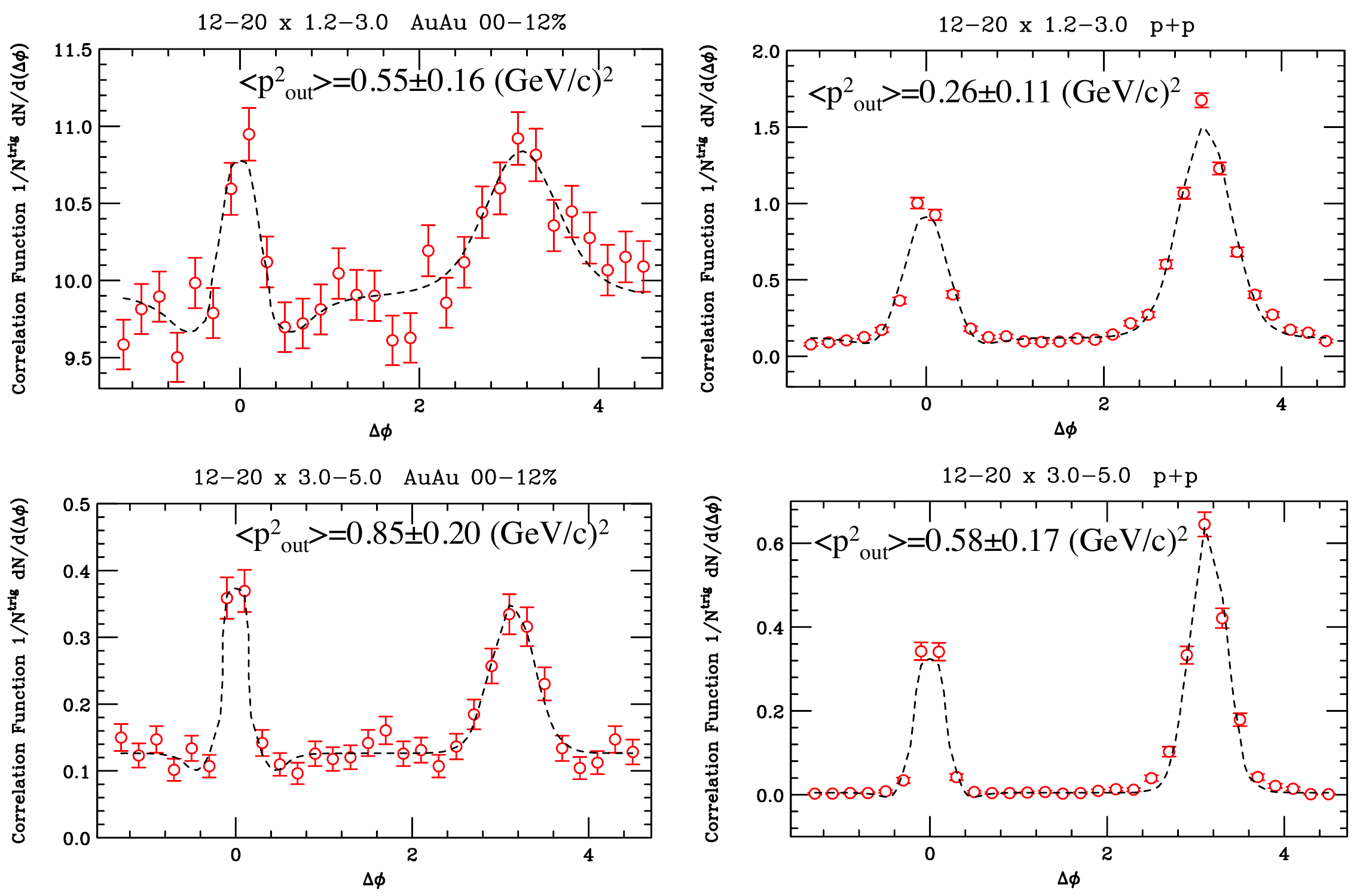}}
\caption[]{\footnotesize Fits to STAR $\pi^0$-hadron correlation functions~\cite{STARPLB760}: Gaussian in $\Delta\phi$ on trigger side  ($\Delta\phi\approx0$), and Gaussian in $p_{\rm out}$ on away side with fitted values of $\mean{p^2_{\rm out}}$ indicated. }   
\label{fig:poutfitstostar}%\vspace*{-1.0pc}
\end{center}
\end{figure}
A recent STAR paper~\cite{STARPLB760} on $\pi^0$-hadron correlations in $\sqsn=200$ GeV Au$+$Au 0-12\% central collisions had very nice correlation functions for large enough $12\leq p_{Tt}\leq 20$ GeV/c so that the $v_2$, $v_3$ modulation of the background was negligible (Fig.~\ref{fig:poutfitstostar}). I made fits to these data to determine $\mean{p^2_{\rm out}}$ so that I could calculate $k_T$ in p$+$p and Au$+$Au using Eq.~\ref{eq:kTcalc}. The results for $3\leq p_{Ta}\leq 5.0$ GeV/c were 
$\sqrt{\mean{k^2_T}}=2.5\pm0.3$ GeV/c for p$+$p and $\sqrt{\mean{k^2_T}}=1.4\pm0.2$ GeV/c, for Au$+$Au, exactly the opposite of azimuthal broadening (Eq.~\ref{eq:broadening}). 
 
After considerable thought, I finally figured out what the problem was and introduced the new $k^{'}_T$.  For a di-jet produced in a hard scattering, the initial $\hat{p}_{Tt}$ and $\hat{p}_{Ta}$ (Fig.~\ref{fig:kTprime}) will both be reduced by energy loss in the medium to become $\hat{p}{'}_{Tt}$ and $\hat{p}{'}_{Ta}$ which will be measured by the di-hadron correlations with $p_{Tt}$ and $p_{Ta}$ in Au$+$Au collisions. The azimuthal angle between the di-jets, determined by the $\mean{k_{T}^2}$ in the original collision, should not change as both jets lose energy unless the medium induces multiple scattering from $\hat{q}$. Thus, without $\hat{q}$ and assuming the same fragmentation transverse momentum $\rms{j_T}$ in the original jets and those that have lost energy, the $p_{\rm out}$ between the away hadron with $p_{Ta}$ and the trigger hadron with  $p_{Tt}$ will not change;  but the $\mean{k{'}_{T}^2}$ will be reduced because the ratio of the away to the trigger jets $\hat{x}{'}_h=\hat{p}{'}_{Ta}/\hat{p}{'}_{Tt}$ will be reduced. Thus the calculation of $k{'}_T$ from the di-hadron p$+$p measurement to compare with Au$+$Au measurements with the same di-hadron $p_{Tt}$ and $p_{Ta}$ must use the values of $\hat{x}_h$, and $\mean{z_T}$ from the Au$+$Au measurement to compensate for the energy lost by the original dijet in p$+$p collisions.

The same values of $\hat{x}_h$, and $\mean{z_t}$ in Au$+$Au and p$+$p simplify Eqs.~\ref{eq:broadening} and \ref{eq:kTcalc} to:
\begin{equation}\mean{\hat{q} L}/2=\left[\frac{\hat{x}_h}{\mean{z_t}}\right]^2_{AA} \;\left[\frac{\mean{p^2_{\rm out}}_{AA} - \mean{p^2_{\rm out}}_{pp}}{x_h^2}\right] \label{eq:coolqhat}\end{equation}
\noindent
from which one could immediately get a reasonable answer for $\mean{\hat{q} L}/2$ from the $\mean{p^2_{\rm out}}$ results indicated on Fig.~\ref{fig:poutfitstostar} if the values of $\hat{x}_h$ and $\mean{z_t}$ in the Au$+$Au measurement are known.

\subsubsection{How to calculate $\mean{\hat{q} L}$ from the Au$+$Au (and p$+$p) measurements}
At RHIC the $\pi^0$ high $p_T$ spectra all have the same $p_T^{-n}$ dependence, $n=8.10\pm 0.05$ in p$+$p and Au$+$Au collisions for all centralities measured at $\sqrt{s}=200$ GeV~\cite{ppg080}. From the Bjorken parent-child relation~\cite{JacobNPB113}, the power $n$ in $p_T^{-n}$ is the same in the jet and fragment ($\pi^0$) $p_T$ spectra.  Also a triggered $\pi^0$ with $p_{Tt}$ is weighted to higher $z_t=p_{Tt}/\hat{p}_{Tt}$, than $\mean{z_t}$ of a fragmentation function because the effective fragmentation function given $p_{Tt}$ is biased by the steeply falling $p_T$ spectrum to $z_t^{n-2}$ times the unbiased fragmentation function. In any case $\mean{z_T}$ can be calculated from the measured $\pi^0$ $p_T$ spectrum~\cite{ppg029}. (For the present discussion, STAR measured $\mean{z_t}=0.80\pm 0.05$ from their p$+$p data~\cite{STARPLB760}.)

\begin{figure}[!h]  
\begin{center}
\raisebox{+0.0pc}{{\footnotesize a)}\hspace*{-1.0pc}\includegraphics[width=0.455\textwidth]{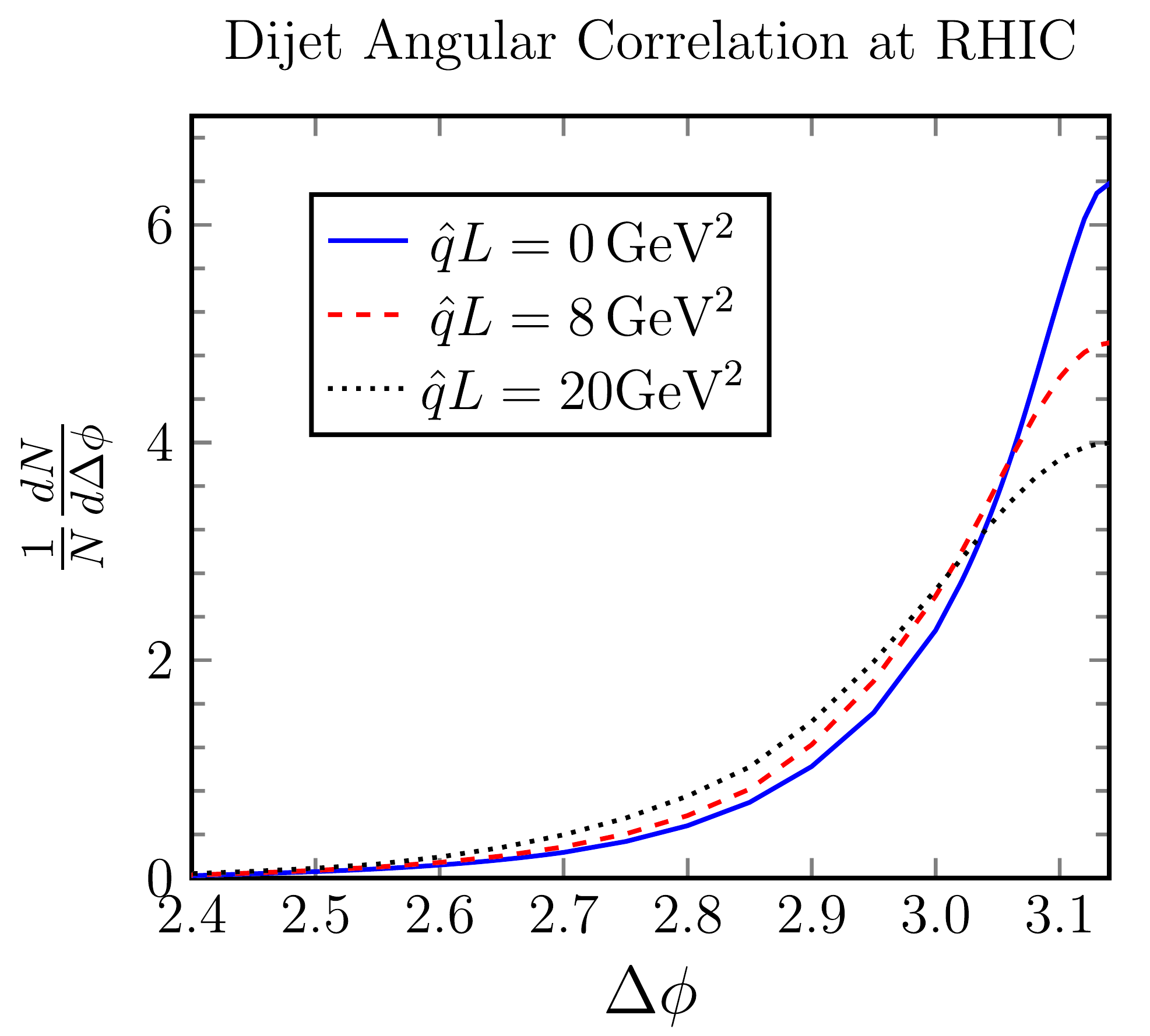}}\hspace*{1.0pc}
\raisebox{+0.0pc}{{\footnotesize b)}\hspace*{-1.0pc}\includegraphics[width=0.45\textwidth]{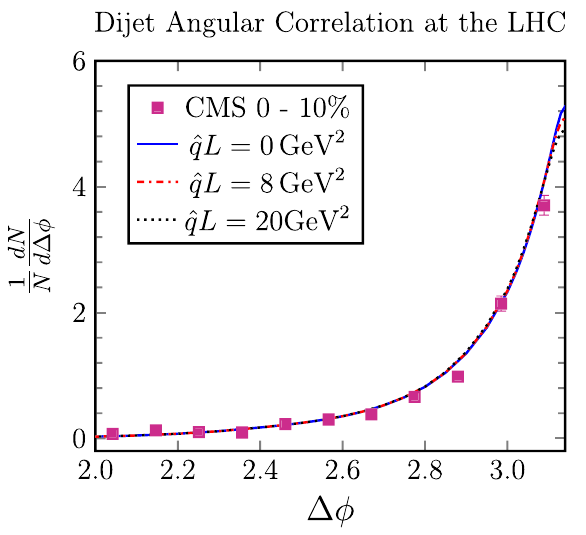}}
\caption[]{\footnotesize Prediction by Al Mueller and collaborators~\cite{MuellerPLB763} of the di-jet azimuthal decorrelation as a function of $\hat{q}L$ for a) 35 GeV jets at RHIC $\sqsn=200$ GeV; and b) 50 GeV jets at the LHC $\sqsn=2.76$ TeV where ``$p_T$ broadening effects are negligible"~\cite{MuellerPLB763}. }   
\label{fig:MuellerqhatL}%\vspace*{-0.12in}
\end{center}
\end{figure}

The ratio of the away jet to the trigger jet transverse momenta $\hat{x}_h=\hat{p}_{Tt}/\hat{p}_{Ta}$ can be calculated from the  away particle $x_h=p_{Ta}/p_{Tt}$ distributions, which were also given in the STAR paper. The formula is~\cite{ppg029}:
    \begin{equation}
\left.{dP \over dp_{Ta}}\right|_{p_{Tt}}  = {N\,(n-1)}{1\over\hat{x}_h} {1\over {(1+ {x_h \over{\hat{x}_h}})^{n}}} \,  
\qquad . \label{eq:jetratio} \end{equation} 
This enabled me to calculate $\mean{\hat{q} L}$ from the $\mean{p^2_{\rm out}}$ results indicated on Fig.~\ref{fig:poutfitstostar}, now with sensible results (Table~\ref{tab:star-PLB760}). The results in the two $p_{Ta}$ bins are at the edge of agreement, different by 2.4$\sigma$, but both are $>2.6 \sigma$ from zero. These results leave several open questions, see Ref.~\cite{MJTPLB771} for the discussion. 
   \begin{table}[!t]\vspace*{-0.0pc} \label{tab:star-PLB760}%\vspace*{-1.0pc}
\begin{center}
\caption[]{Tabulations for $\hat{q}$--STAR $\pi^0$-h~PLB760(2016)689--696-MJT PLB(2017)}%\vspace*{-1.0pc} 
{\begin{tabular}{cccccc}  
%\hline
%STAR PLB760(2016)689--696\\
\hline
\hline
$\sqsn=200$GeV &$\mean{p_{Tt}}$&$\mean{p_{Ta}}$&$\sqrt{\mean{k_{T}^2}}_{AA}$& $\sqrt{\mean{k{'}_{T}^2}}_{pp}$ & $\mean{\hat{q} L}$\\
 \hline
Reaction & GeV/c & GeV/c &GeV/c & GeV/c &GeV$^2$ \\
 \hline
Au$+$Au 0-12\%&14.71&1.72&$2.28\pm0.35$&$1.01\pm 0.18$& $8.41\pm2.66$\\
\hline
Au$+$Au 0-12\%&14.71 &3.75&$1.42\pm0.22$ &$1.08\pm 0.18$&$1.71\pm 0.67$\\
\hline\\[-1.0pc] 
\hline
\end{tabular}} \label{tab:star-PLB760}%\vspace*{-1.0pc}
\end{center}\vspace*{-0.1pc}
\end{table}
However there is a nice prediction of $\Delta\phi$ for for 35 GeV Jets at RHIC~\cite{MuellerPLB763} for several values of $\mean{\hat{q} L}$ (Fig.~\ref{fig:MuellerqhatL}). An amusing test would be to see if the present method gives the same answers for $\mean{\hat{q} L}$ by calculating $\mean{p^2_{\rm out}}$ of the predictions.
\end{document}